\documentclass[aps,prl,twocolumn,notitlepage,superscriptaddress,footinbib,amsmath,amssymb,floatfix]{revtex4-1}
\usepackage{graphicx,mathrsfs,amsfonts,braket,bm,multirow,setspace}
\usepackage{siunitx}
\usepackage[usenames,dvipsnames]{color}
\usepackage{hyperref}
\hypersetup{colorlinks,
  citecolor=Blue,
  linkcolor=Blue,
  urlcolor=Blue
}

\renewcommand{\vec}{\bm}

\begin{document}

\title{Observation of a Transition Between Dynamical Phases in a Quantum Degenerate Fermi Gas}

\author{Scott Smale}
\altaffiliation{These authors contributed equally to this work.}
\affiliation{Department of Physics, University of Toronto, Ontario M5S~1A7, Canada}
\author{Peiru He}
\altaffiliation{These authors contributed equally to this work.}
\affiliation{JILA, National Institute of Standards and Technology, and University of Colorado,\\ 
Department of Physics, University of Colorado, Boulder, CO 80309, USA}
\affiliation{Center for Theory of Quantum Matter, University of Colorado, Boulder, Colorado 80309, USA}
\author{Ben A. Olsen}
\altaffiliation{current address: Yale-NUS College, 138527 Singapore}
\author{Kenneth G. Jackson}
\author{Haille Sharum}
\author{Stefan Trotzky}
\affiliation{Department of Physics, University of Toronto, Ontario M5S~1A7, Canada}
\author{Jamir Marino}
\altaffiliation{current address: Harvard University, Dept.\ of Physics, Cambridge, MA 02138, USA}
\author{Ana Maria Rey}
\altaffiliation{Corresponding author emails: arey@jilau1.colorado.edu, jht@physics.utoronto.ca}
\affiliation{JILA, National Institute of Standards and Technology, and University of Colorado,\\
Department of Physics, University of Colorado, Boulder, CO 80309, USA}
\affiliation{Center for Theory of Quantum Matter, University of Colorado, Boulder, Colorado 80309, USA}
\author{Joseph H. Thywissen}
\altaffiliation{Corresponding author emails: arey@jilau1.colorado.edu, jht@physics.utoronto.ca}
\affiliation{Department of Physics, University of Toronto, Ontario M5S~1A7, Canada}
\affiliation{Canadian Institute for Advanced Research, Toronto, Ontario M5G~1Z8, Canada}

\begin{abstract}
A proposed paradigm for out-of-equilibrium quantum systems is that an analogue of quantum phase transitions exists between parameter regimes of qualitatively distinct time-dependent behavior. 
Here, we present evidence of such a transition between dynamical phases in a cold-atom quantum simulator of the collective Heisenberg model. 
Our simulator encodes spin in the hyperfine states of ultracold fermionic potassium. Atoms are pinned in a network of single-particle modes, whose spatial extent emulates the long-range interactions of traditional quantum magnets. 
We find that below a critical interaction strength, magnetization of an initially polarized fermionic gas decays quickly, while above the transition point, the magnetization becomes long-lived, due to an energy gap that protects against dephasing by the inhomogeneous axial field. 
Our quantum simulation reveals a non-equilibrium transition predicted to exist but not yet directly observed in quenched s-wave superconductors.
\end{abstract}

\maketitle

\smallskip
\noindent\textbf{{Introduction}}

\noindent
The challenge faced in understanding out-of-equilibrium systems is that the powerful formalism of statistical physics, which has allowed a classification of quantum phases of matter based on simple principles such as minimization of free energy, does not apply. 
A diverse range of non-equilibrium phenomena have been observed, including synchronization \cite{Deutsch:2010ky,Solaro:2016iv,Piechon:2009cr,Norcia2017}, self-organization \cite{Baumann2010,Klinder2015,Leonard2017,Li:2017}, quantum chaos \cite{Neill2016,Chaudhury}, Loschmidt echo singularities \cite{Jurcevic2017}, and time crystals \cite{Zhang:2017ci,Choi2017}. 
A proposed organizing principle is that transitions, reminiscent of those found between thermodynamic ground states, can also be found between dynamical phases \cite{eckstein09,Schiro10,Sciolla,gambassi,zhang}. 

In general terms, a non-equilibrium phase transition is characterized by the existence of a critical point separating phases with distinct dynamical properties in many-body systems. 
The analogue of thermodynamic order parameters is found in long-time-average observables, which have a non-analytic dependence on system parameters. 
In driven open systems, energy and particle number are not necessarily conserved, and non-equilibrium transitions are typically signaled by different steady states that occur upon varying system parameters such as pump or loss rates \cite{Zoller,Altman,Marino}, independently of initial conditions. In closed systems, dynamics are often initiated by quenching control parameters, with qualitatively distinct behaviors observed below, above, or at a critical point \cite{eckstein09,Schiro10,Sciolla,gambassi} that can depend on the initial state of the system. 
The label ``dynamical phase transition'' has been applied not only to the boundary between two dynamical phases, but also to the non-analytic behavior in real-time dynamics of the return probability-amplitude \cite{Flaschner:2018ch,Jurcevic2017}, which does not require an order parameter to be defined \cite{heyl13}. The phenomenon under investigation in our work is the former case, which we will refer to as a {\em transition between dynamical phases} (TDP) to avoid confusion. 
The theoretical study of such transitions have encompassed a broad range of platforms including collective spin models \cite{BoyanMF, Bojan2016b, PhysRevB.99.045128}, non-equilibrium phases of superconductors \cite{Dukelsky2004,Barank04,Gurarie2015}, interacting fermions and bosons on the lattice~\cite{kollath, eckstein09,Schiro10,Sciolla,PhysRevE.93.032219}, and quantum field theories~\cite{gambassi, marino2}; however, experimental investigations have so far been restricted to self-trapping transitions in bosonic systems \cite{Anker2005,Albiez2005,Levy2007,Abbarchi2013,Reinhard2013} and the transverse-field Ising model realized with trapped-ion chains \cite{zhang}.

Here, we report the observation of a transition between two dynamical phases of a quantum degenerate Fermi gas. The sample under investigation consists of neutral potassium atoms ($^{40}$K) confined in a harmonic optical trap and cooled to nanokelvin temperatures. 
The controllable interactions of this closed quantum system enable a broad search for non-equilibrium phenomena that arise from the interplay of atomic contact interactions, quantum statistics, and motion. Using collective magnetization as an order parameter, system dynamics are observed directly, and compared to theoretical models at various levels of approximation. 

\begin{figure}[tb!]
\centering
{\includegraphics[width=\columnwidth]{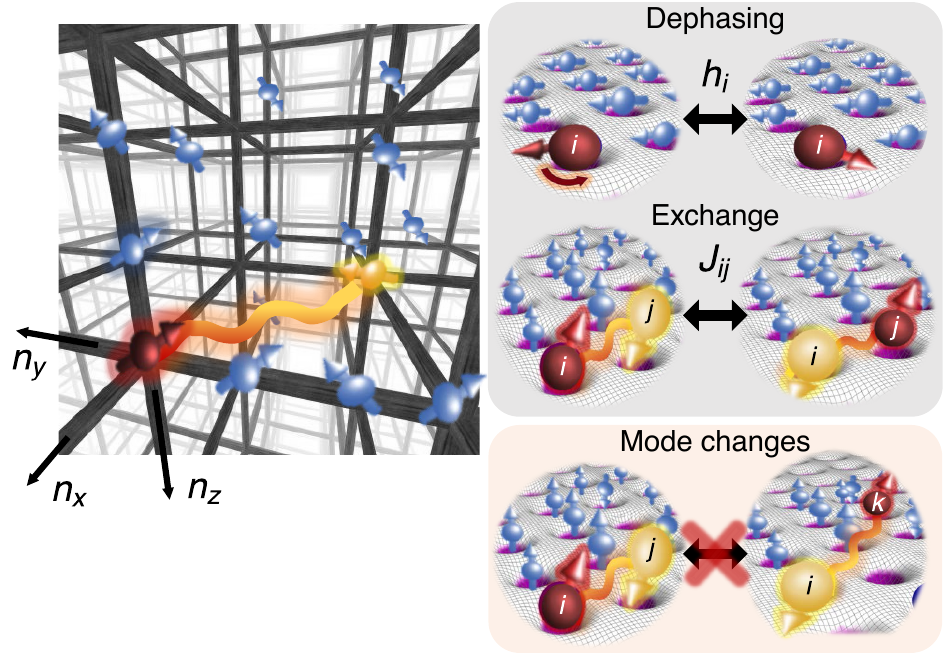}}
{\caption{{\bf Simulation of the collective Heisenberg model with a local inhomogeneous axial field using weakly interacting fermions in a mode space lattice.}
Each site in mode space (left) has an occupancy of 0 or 1 and experiences a local field $h_i$ that causes single spin precession at a rate that depends on the mode (top right). Atoms experience long-range spin-exchange interactions $J_{ij}$ (middle right). Mode-changing collisions would move atoms between sites in mode space (bottom right), but are not included in the spin model.}
\label{fig:model}}
\end{figure}

We understand and analyze our system through a mapping of the single-particle eigenstates of the harmonic trap onto a lattice in mode space, as depicted in Fig.~\ref{fig:model}. By tuning the interaction strength to suppress collisions that would change the occupancy of the modes, the atoms become pinned on the conceptual lattice, enabling the description of our system with a spin model~\cite{Martin:2013kl,Koller:2015dg,Koller:2016el}, here
\begin{equation} \label{eq:H}
\hat{H}/\hbar = \sum_i h_i \hat{s}^Z_i - \sum_{i,j} J_{ij} \bm{\hat{s}}_i \cdot \bm{\hat{s}}_j,
\end{equation}
where $\bm{\hat{s}}_i = 1/2\{\hat{\sigma}_i^X,\hat{\sigma}_i^Y,\hat{\sigma}_i^Z\}$ are spin-$1/2$ operators acting on the $i$th atom, and $X$, $Y$, and $Z$ denote orientations in Bloch space. This is the collective Heisenberg model (CHM), a canonical model for magnetism \cite{Auerbach1994}, in which the non-local spin-spin couplings $J_{ij}$ compete with an inhomogeneous axial field, $h_i$ \cite{Martin:2013kl,Koller:2015dg,Koller:2016el}.
Similar treatments of fermionic systems using a spin model have been employed successfully in optical lattice clocks \cite{Martin:2013kl,Zhang2014,Rey2014,Swallows:2010er,2018bromley} in the microkelvin regime. However, in those experiments, undesirable inelastic collisions limited the number, $N \lesssim 50$, and prevented the study of transitions between well defined dynamical phases. The stability of low-lying hyperfine states and control over interactions in our experiment allows us to explore many-body dynamics in macroscopic ultracold samples of $N \simeq 3\times 10^4$ alkali atoms at nanokelvin temperatures and test the spin model in this new regime. 

We find that below a critical interaction strength, the magnetization of an initially polarized gas quickly decays, while above this critical point, the magnetization becomes long-lived, and is protected by an energy gap against inhomogeneous field-induced dephasing. These observed dynamical phases are a manifestation of an emergent property in a many-body dynamical quantum system. The transition would be absent for small particle number, since the emergence of a sufficiently strong gap from weak two-body collisions relies upon a collective nonlinearity. 
We then implement a many-body echo sequence, which is a direct test of reversibility. This allows us to identify the boundaries of the parameter regime in which the complex far-from-equilibrium dynamics of interacting fermions is quantitatively described by the CHM. 
The successful mapping allows us to implement quantum simulations of the non-equilibrium phases predicted to exist in quenched s-wave superconductors by Richardson-Gaudin models \cite{Dukelsky2004,Barank04,Gurarie2015} but not yet directly observed, given the need for ultra-fast probes \cite{Matsunaga1145}.


\noindent\textbf{{Results}}

\noindent
The simulation cycle begins with a non-interacting sample fully polarized in the lower spin state $\ket{\downarrow}$, which ensures that no site in the mode lattice is doubly occupied. Non-equilibrium dynamics are initiated by a fast radio-frequency (rf) pulse that rotates the collective magnetization into the $XY$ plane. 
The time evolution of transverse magnetization is probed using a Ramsey sequence: following the initial $\pi/2$ pulse, atoms evolve for a variable time $t$, after which a second $\pi/2$ pulse is applied, and the total populations in the $\ket{\downarrow}$ and $\ket{\uparrow}$ states are measured with a Stern-Gerlach technique. Shot-to-shot field drifts on the microtesla scale prevent a reproducible accumulated phase in the Ramsey sequence. We estimate the magnitude of the transverse coherence by repeating the sequence at least 10 times and using a maximum-likelihood estimator that assumes a randomized interferometric phase (see Methods and Materials for further details). This procedure measures the total transverse magnetization $ 2\sqrt{({\mathcal S}^X)^2+({\mathcal S}^Y)^2}/N$, where ${\mathcal S}^{X,Y,Z}=\langle {\hat S}^{X,Y,Z}\rangle$ with ${\hat S}^{X,Y,Z}= \sum_i {\hat s}_i^{X,Y,Z}$ as collective spin operators. However, since ${\mathcal{S}}^Z$ is a constant of motion in our simulation, and set to zero by the first $\pi/2$ pulse, we can simply interpret the signal as $2 {\mathcal S}/N$, with ${\mathcal S}=\sqrt{\sum_{p =X,Y,Z}{(\mathcal S}^p)^2}$ the total magnetization.

The optical confinement creates a potential that is approximately harmonic, with frequencies $\bm{\omega}=\{ \omega_x, \omega_{y}, \omega_z \} = 2 \pi \times \{ 395, 1140, 950\}$\,Hz along three spatial directions.  
Spin-dependent curvature in the confinement potential produces a further shift $\pm \bm{\Delta\omega}$ in the oscillator frequency between the $\ket{\downarrow}$ and $\ket{\uparrow}$ states. Since the resultant energy shift depends linearly on the index of the single-particle motional eigenstates, labelled by $\bm{n}_i=\{ n_{i}^{x}$, $n_{i}^{y}$, $n_{i}^{z} \}$, 
it constitutes an inhomogeneous axial field in mode space, $h_i = 2 \bm n_i \cdot \bm {\Delta\omega}= 2(n_{i}^{x} \Delta\omega_x + n_{i}^{y} \Delta\omega_y + n_{i}^{z} \Delta\omega_z)$. The strength of the inhomogeneity is tuned in two ways: using the polarization of one of the laser beams forming the optical trap to change $\bm {\Delta\omega}$ (see Methods and Materials), and using temperature to change the average mode index $\bar{n}$ within the range $20$ to $30$. 

Interactions are proportional to the s-wave scattering length $a$ of the colliding atoms, which are tuned by a magnetic Feshbach resonance near \SI{20}{\milli\tesla} \cite{RevModPhys2010}. 
We factor $J_{ij}$ as $U \mathcal{J}_{ij}$, where $U=4\pi a \sqrt{m \omega_x \omega_y \omega_z /\hbar }$ sets an overall scale, and ${\mathcal J}_{ij}$ is a mode-dependent coupling factor proportional to the density-density overlap of the single-particle eigenmodes of the $i$th and $j$th particles (see Supplementary Materials). Due to the extended nature of the motional wave functions, the $\mathcal J_{ij}$ are long-ranged, $\sim 1/\sqrt{|n_i^p - n_j^p|}$ in each direction $p=x,y,z$. 
The TDP is observed in a weak-scattering regime, where atoms remain frozen in their initial modes, and dynamics involve only spin degrees of freedom \cite{Piechon:2009cr,Natu:2009gt,Du:2009hm}. 
For these experiments, we needed to improve by an order of magnitude the accuracy to which the zero-crossing field  $B_\mathrm{zc}$ (at which $a=0$) was known; data used to determine $B_\mathrm{zc}=20.907(1)$\,mT are presented in the Supplementary Material. 

Figure \ref{fig:phasediagram} shows an exploration of the non-equilibrium phase diagram using total magnetization at 100~ms, ${\mathcal S}(t=100\,\mathrm{ms})$, as the order parameter. Simulations were run with scans of mean interaction strength $J = \sum_{i,j} J_{ij}/N^{2}$ (Fig.~\ref{fig:phasediagram}A,B,C) or scans of axial field inhomogeneity $\widetilde{h}= \sqrt{\sum_i h_i^2/N - (\sum_i h_i/N)^2}$ (Fig.~\ref{fig:phasediagram}E,F,G). 
Several distinct regions appear: a dynamical ferromagnet with high persistent $\mathcal{S}$ at large positive or negative $NJ$, and a paramagnetic phase with  low $\mathcal{S}$ at smaller $|NJ|$.

\begin{figure*}[tb!]
\centering{\includegraphics[width=\textwidth]{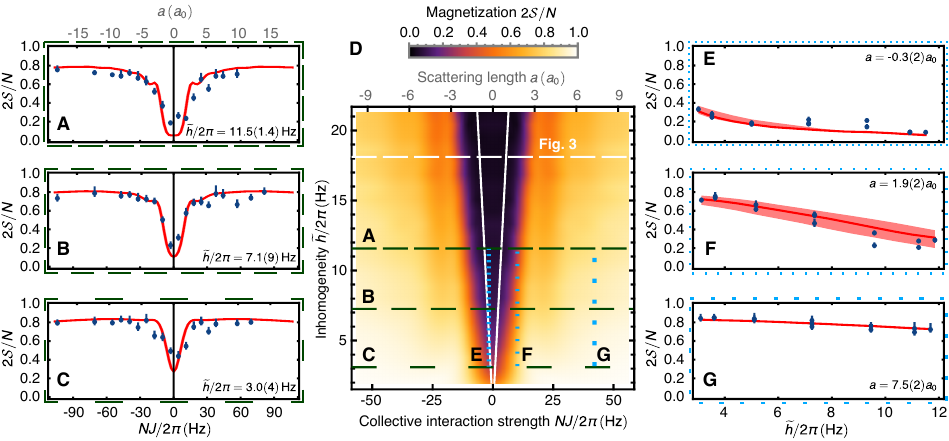}}
\caption{ {\bf Non-equilibrium phase diagram.} (\textbf{A}--\textbf{C},\textbf{E}--\textbf{G}) show magnetization $\mathcal{S}$ at $t=100$\,ms versus interaction strength $J$ and inhomogeneity $\widetilde{h}$. Blue circles are experimental data, and red lines are theory calculations scaled by $e^{-\Gamma(a) t}$ at $t=100$\,ms (see Fig.~\ref{fig:reverse}). These are cuts through a non-equilibrium phase diagram (\textbf{D}), calculated using magnetization at $t=100$ as the order parameter. The dashed white line shows a cut at the inhomogeneity used in Fig.~\ref{fig:Transition}. The solid white lines show the steady-state ($t\rightarrow\infty$) phase boundary for $\alpha/\beta=0.27$, determined by the critical point of the magnetization in cuts of constant $\widetilde{h}$. Error bars on data points are statistical; bands in theory correspond to uncertainty in $B_\mathrm{zc}$. All theory calculations except for the solid white line are finite time thermally averaged numerics as described in the main text.
\label{fig:phasediagram}}
\end{figure*}

Measurements are compared to a mean-field treatment of Eq.~\ref{eq:H}, where the $j$th atom experiences an effective magnetic field, $\bm{B}_j$, which depends only on the local field, $h_j$, and the average magnetization of the other atoms in the ensemble,
\begin{equation} \label{eq:Hmf}
\hat{H}_\mathrm{mf}/\hbar=\sum_j \bm{\hat s}_j \cdot \bm{B}_j
\end{equation}
where $\bm{B}_j =h_j \vec{Z} - \sum_{i} 2 J_{ij} \langle \bm{\hat s}_i \rangle$.
Here the indices $i,j$ run over a set of $N$ populated modes drawn from a finite-temperature Fermi-Dirac distribution, and $\vec Z$ is a unit vector. 
The transition between dynamical phases is a consequence of the opening of an interaction-energy gap between the fully polarized manifold and the remainder of the Hilbert space, as further discussed below. 
Numerical solutions of the corresponding $3N$ non-linear Bloch equations (see Supplementary Materials) show that above some critical interaction strength, $J_c(\widetilde{h})$, the dynamics become gapped and ferromagnetic order (meta)stabilizes, while below $J_c(\widetilde{h})$, the gas partially demagnetizes. 
In the ungapped phase, exchange interactions are not strong enough to prevent demagnetization, rendering the system prone to dephasing induced by inhomogeneous $h_i$.
In contrast to thermodynamic ferromagnetism, which occurs only for $J>0$, the dynamical ferromagnet is a spin-locked state that can be stabilized by either repulsive or attractive exchange interactions.

The red lines in Fig.~\ref{fig:phasediagram} show ${\mathcal S}(t=100\,\mathrm{ms})$ calculated by this model, using ab-initio determination of $J$, a set of field curvatures that match all observations in this manuscript (see Supplementary Materials), and a decay envelope, $e^{-\Gamma(a) t}$, discussed in detail further below. The white solid line in Fig.~\ref{fig:phasediagram}D shows the TDP steady-state $(t\rightarrow \infty)$ phase boundary which sets $J_c(\widetilde{h})$. 
While we are unable to probe the system at infinite time and thereby reveal the strict steady-state limit we find, as shown by the corroboration in Fig.~\ref{fig:phasediagram}, that 
$t=100$\,ms $\approx 40\,(\omega_x/2\pi)^{-1}$ 
is sufficiently long to capture the non-equilibrium phase diagram as a function of $J$ and $\widetilde{h}$.

To gain understanding of the scaling expected near the TDP one can use a simplified ``all-to-all'' model, in which coupling constants are replaced by their mean value, $J_{ij}\to J$. In this limit, Eq.~\ref{eq:Hmf} becomes integrable and maps to the Bardeen-Cooper-Schrieffer Hamiltonian for fermionic superconductors expressed in terms of Anderson pseudo-spin \cite{Anderson1958}. Borrowing the methodology developed for dealing with dynamical phases in superconductors \cite{Yuz, Yuz2,Gurarie2015}, one can obtain the frequency spectrum ruling the non-equilibrium dynamics from the roots of ${L}^2(u)$, where $\vec{L}(u)$ is the \emph{Lax vector} of the auxiliary variable $u$ (see Supplementary Materials). The roots can be found using the property that ${L}^2(u)$ is an integral of motion of the dynamics, and can be evaluated for convenience at time $t=0$. 
When the roots compress in the neighborhood of the real axis, the long-time limit of $\mathcal{S}(t)$ relaxes to a zero. This corresponds to the normal phase, or ``phase I'' in the language of superconductors \cite{Gurarie2015}. On the other hand, the appearance of a pair of complex conjugate roots determines the TDP critical point and the emergence of a phase characterized by a non-zero steady-state order parameter, $\mathcal{S}(\infty)>0$, i.e.\ ``phase II''. While the precise scaling of the order parameter near the TDP can be complex, since it is determined by the spectrum of the $h_i$, we find that above the critical point in our system it can be approximated by the analytic expression
\begin{equation} \label{eq:Ssss}
\mathcal{S}(\infty) \approx \frac{\sqrt{3}\alpha\,\widetilde{h}}{2J_{\rm eff}}\cot\left(\frac{\sqrt{3}\alpha\,\widetilde{h}}{N J_{\rm eff}}\right)\,.
\end{equation}
This formula is exact (with $\alpha=1$ and $J_{\rm eff}=J$) for the case of a one-dimensional system $\Delta\omega_{y,z}=0$ at zero temperature. To account for non-collective interactions, higher dimensions, and finite temperature, we introduce renormalization parameters $\alpha$ and $J_{\rm eff}=\beta J$.
The critical interaction strength is $J_c(\widetilde{h})= 2\sqrt{3}\alpha \widetilde{h}/(\beta N\pi)$.
An alternate order parameter in the non-equilibrium ferromagnetic phase is the gap frequency:
\begin{equation} \label{eq:Omega}
\Omega = 2 |J_{\rm eff}| \mathcal{S}(\infty)\approx \sqrt{3} \alpha\,\widetilde{h}\cot\left(\frac{\sqrt{3}\alpha\,\widetilde{h}}{NJ_{\rm eff}}\right).
\end{equation} In the ferromagnetic phase, $\mathcal{S}(t)$ exhibits transient oscillations at the gap frequency $\Omega$, which slowly damp as it reaches $\mathcal{S}(\infty)$. The gap frequency goes to zero at $J_c( \widetilde{h})$ in a non-analytic manner. As discussed below, we observe each of these signatures in the quantum simulation.

The collective nature of these phenomena is emphasized by an alternative interpretation of the gap. The initial $\pi/2$ pulse can be said to create a superposition of $\ket{S=N/2, m_S}$ Dicke states, where $S$ and $m_S$ are eigenvalues of the collective operators ${\hat{S}}^2=\sum_{p=X,Y,Z}(\hat{S}^p)^2 $ and $\hat{S}^Z$ respectively. All $m_S$ states have the same energy in the rotating frame of the rf pulse. A finite energy gap $\hbar \Omega$ inhibits the production of spin waves (generated by the inhomogeneous $h_i$) and keeps the dynamics within the collective Dicke manifold: flipping a single spin would reduce $S$ to $(N/2-1)$, and change the exchange energy, proportional to $J \hat{S}^2$, by $NJ$.

\begin{figure*}[t!]
\centering{\includegraphics[width=0.9\textwidth]{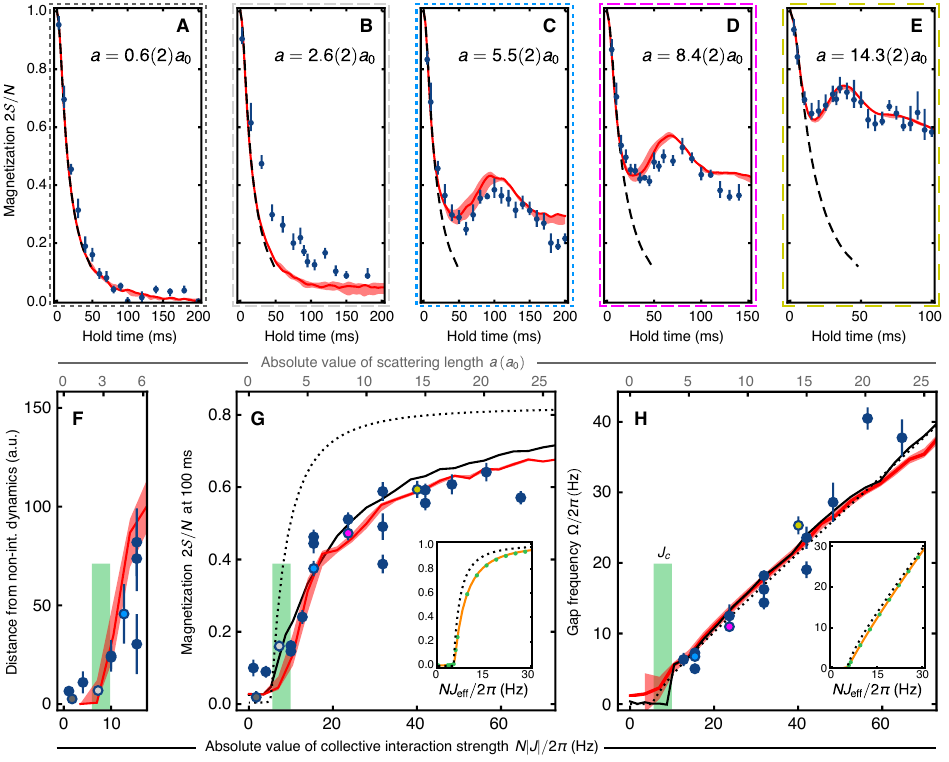}
\caption{{\bf  Transition between dynamical phases}.
(\textbf{A}-\textbf{E}) Time-dependent magnetization $\mathcal{S}(t)$ for fixed $\widetilde h = 2 \pi \times 18.1(1)$\,Hz. As interaction strength increases, data (blue points, with statistical error bars) and theory (red lines, with error bands due to bias field uncertainty) deviate from non-interacting dynamics (dashed black curve) after some time. 
(\textbf{F}) Experiment and theory are compared via $\chi^2$ distance to non-interacting dynamics, both significantly deviating for $NJ/2\pi\gtrsim10$\,Hz, or $a\gtrsim 3a_0$.
(\textbf{G}) The interpolated magnetization at 100\,ms agrees well with numerical solutions of the mean-field dynamics with thermal averaging at finite time (red band) and without thermal averaging in steady state (solid black line), but only qualitatively with the analytic approximation of Eq.~\ref{eq:Ssss} (dotted black line).
(\textbf{H}) The gap frequency $\Omega$ shows good agreement between data and all three levels of theory. $\Omega$ is fit to the analytic formula of Eq.~\ref{eq:Omega} to find $N J_c/2\pi=7.8(1.1)$\,Hz, indicated by the green band. 
\textbf{Insets in G, H} compare the analytic approximation of Eq.~\ref{eq:Ssss} and Eq.~\ref{eq:Omega} 
(dotted black line) with the exact all-to-all solution from the Lax vector approach (orange line) and from the corresponding mean-field numerical solution (green circles). 
Apart from insets, theory curves are scaled by $e^{-\Gamma(a) t}$ to account for beyond-spin-model processes (see Fig.~\ref{fig:reverse}). 
Error bars on data points are statistical; the theory uncertainty (red band) is dominated by $B_\mathrm{zc}$, and the width of the green $J_c$ band is $\pm$ two standard deviations.
}
\label{fig:Transition}}
\end{figure*}

Figures~\ref{fig:Transition}A-E show the qualitative change in dynamical behavior as $J$ crosses $J_c$ for fixed $\widetilde{h}$.
Below the TDP, $\mathcal{S}(t)$ decays monotonically in time (Fig.~\ref{fig:Transition}A), but above the transition, $\mathcal{S}(t)$ oscillates around a non-zero magnetization (Fig.~\ref{fig:Transition}C,D,E). 
All observations can be reproduced by the same theoretical model shown in Fig.~2 (red lines), if $\mathcal{J}_{ij}$ are scaled by $0.8$ from their ab-initio values, perhaps due to an increased sampling of trap anharmonicity due to the higher temperature used in this data set to increase $\widetilde{h}$, or due to a renormalization of coupling constants due to resonant mode-changing processes \cite{Koller:2015dg,Koller:2016el}. The TDP is seen in three observables (Figs.~\ref{fig:Transition}F,G,H): first, by a departure from ungapped dynamics; second, by looking for a jump in $\mathcal{S}$ at 100\,ms; and third, by a finite value of $\Omega$.

The $\chi^2$ measure in Fig.~\ref{fig:Transition}F compares both data and calculations to $\mathcal{S}_{J=0}(t)$, the calculated time evolution for $J=0$. The sharp increase near $NJ/2\pi\approx10$\,Hz in both experiment and theory indicates the qualitative deviation of the dynamics from the paramagnetic phase. 
The magnetization at $t=100$\,ms (in Fig.~\ref{fig:Transition}G) also shows an increase near $NJ/2\pi\approx10$\,Hz. 
Numerical solutions of the mean-field dynamics with thermal averaging at finite time (red band) and without thermal averaging in steady state (solid black line) agree well with the experimental data. The simplified all-to-all model (Eq.~\ref{eq:Ssss}, dotted black line) agrees only qualitatively. 
The gap frequency $\Omega$ in Fig.~\ref{fig:Transition}H is found from a fitting function that uses a damped sinusoid for later times, and $\mathcal{S}_{J=0}$ for early times (see Supplementary Materials). 
By fitting the gap parameter to the analytic formula Eq.~\ref{eq:Omega}, we extract a nonzero critical interaction strength $N J_c/2\pi=7.8(1.1)$\,Hz. Using the location of the step in $\chi^2$ (Fig.~\ref{fig:Transition}G), we exclude time sequences with $|a|<3a_0$, where the oscillation frequency diverges. The excellent agreement of all three measures with theory based on Eq.~\ref{eq:H} confirm that our quantum simulator probes the TDP in the collective Heisenberg model.

The significance of the observation is further clarified in Figs.~\ref{fig:Transition}G,H by comparison to various approximation levels. 
Finite-time effects and thermal averaging play a minor role, validating our interpretation of $\mathcal{S}$ at sufficiently large $t$ as the steady-state order parameter. Inhomogeneous coupling ($J_{ij} \neq J$) plays a significant role for $\mathcal{S}$, but less so for $\Omega$. Comparisons to the exact Lax vector analysis (insets to Fig.~\ref{fig:Transition}G,H) show the close similarity of the observed TDP to the phase-I-to-phase-II transition in dynamical superconductors \cite{Dukelsky2004,Barank04,Gurarie2015}.

\begin{figure*}[tb!]
\centering {\includegraphics[width=0.9\textwidth]{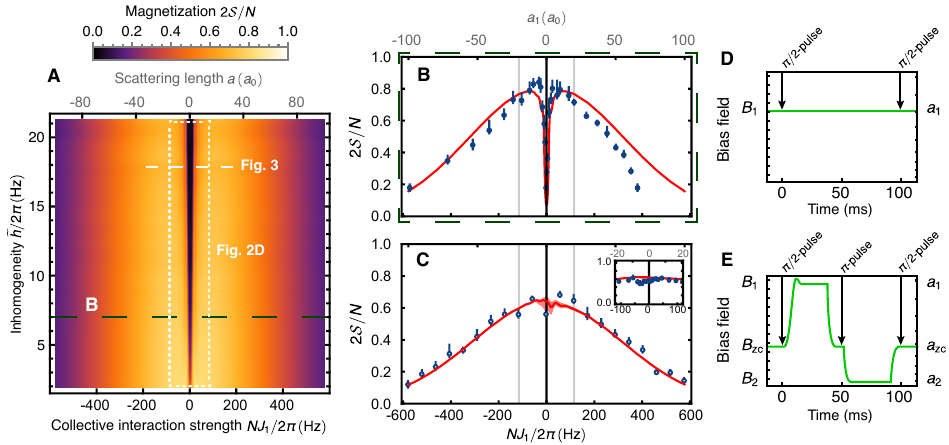}}
\caption{
{\bf From reversibility to the breakdown of CHM simulation.}
(\textbf{A}) Spin-model mean-field dynamical phase diagram across a wider range than Fig.~2D, showing significant dephasing. 
(\textbf{B,D}) Magnetization versus interaction strength, with constant bias field. Here, $\mathcal S$ decreases for very small or very large interactions. The large-$|NJ|$ behavior is captured by augmenting the spin model dynamics with a phenomenological dephasing term (red curves). 
(\textbf{C,E}) Magnetization versus interaction strength, with a many-body echo sequence that reverses the sign of $\hat H \Psi$ at $t/2$. The bias field is held at $B_1$ in the first half of the evolution time, yielding scattering length $a_1$ and collective interaction strength $NJ_1$, while in the second half, $B_2$ is chosen such that $a_2=-a_1$, and $NJ_2=-NJ_1$. 
The initialization, spin-reversal, and readout pulses are performed at $B_\mathrm{zc}$. The magnetization can be recovered for $|NJ|/2\pi\lesssim 100$\,Hz, a region shown in more detail by the Inset of \textbf{C}.
A small systematic error $\Delta B_\mathrm{zc}\sim\SI{2}{\micro\tesla}$ leads to a shift $\Delta a\sim0.3a_0$ in $a_\mathrm{zc},a_1,a_2$, included in the theory curves. 
\label{fig:reverse} }
\end{figure*}

Figure \ref{fig:reverse} describes a further set of simulations that probe the limits of validity in which our system is described by a spin-lattice model. Figures~\ref{fig:reverse}A,B show that $\mathcal{S}(t=100\text{\,ms})$ decreases at sufficiently large $a$, despite a larger gap. This is accompanied by a breakdown in microscopic reversibility, as seen by comparing to a sequence with a many-body reversal of the spin model  
(Fig.~\ref{fig:reverse}C), in which $\hat H \Psi \to - \hat H \Psi$ \cite{Widera2008,Du:2009hm}. 
Signatures of time reversibility within the window $|a| \lesssim 20a_0$ are seen from the nearly $J$-independent dynamics of $\mathcal{S}$ in both the gapped and ungapped phases. 
The reversibility of demagnetization in our system in this regime 
is a significant validation of Eq.~\ref{eq:H}, since the many-body echo sequence does not reverse all terms (e.g., the spin independent harmonic oscillator term) in the full Hamiltonian. 

Two processes prevent full reversibility in our simulation: stray magnetic field gradients and collisional processes. These are quantified by introducing an empirical dephasing rate $\Gamma(a) = \Gamma_0+ (a/a_0)^{2} \gamma$ to the transverse magnetization, such that $\mathcal{S}^{X,Y}(t)\to \mathcal{S}^{X,Y}(t) e^{-\Gamma(a) t}$. 
Figures~\ref{fig:reverse}B,C compare data with calculations using a best-fit $\Gamma_0^{-1}=0.57$\,s without echo and $\Gamma_0^{-1}=0.25$\,s with echo, and $\gamma^{-1} = 600$\,s. 
Here $\Gamma_0$ accounts for the single-particle mode-changing processes generated by magnetic field gradients, which are enhanced by a spin-reversal \cite{Deutsch:2010ky,Koller:2015dg}.
$\gamma$ parametrizes mode-changing collisions that take place at a rate that increases quadratically with $a$, and takes a value anticipated by kinetic theory for our experimental density, temperature, and polarization (see Supplementary Materials). At larger $|a|$, coherence is lost, and a quantum picture becomes unnecessary, as shown by the success of a semi-classical picture to describe diffusive transverse demagnetization \cite{Lhuillier:1982wc,Piechon:2009cr,Johnson,Gully,McGuirk2002,Kohl2013}.
Figure~\ref{fig:reverse}A shows an extended phase diagram, where it can be seen that trapped fermions simulate the CHM only in a restricted parameter window of many-body quantum coherence.

\noindent\textbf{{Discussion}}

\noindent
In sum, we demonstrate the existence of a TDP in a neutral Fermi gas in a regime of reversible dynamics near the zero-crossing of a Feshbach resonance. We outline a direct connection to non-equilibrium phases in the Richardson-Gaudin models for superconductivity, thereby extending
experimental observations of TDP's beyond prior manifestations in Josephson and Ising-type systems \cite{Anker2005,Albiez2005,Levy2007,Abbarchi2013,Reinhard2013,zhang}. 
Moreover, the  excellent agreement between spin-model calculations and a two-axis exploration of the dynamical phase diagram with $> 10^4$ spins provide experimental evidence of the scaling behavior and universal character of TDPs.

The collective nature of the dynamics observed here could protect many-body states in other systems of interest for applied quantum technologies. For example, gap protection would increase the coherence time in optical lattice clocks operated in the quantum degenerate regime \cite{Campbell}. 
Furthermore, using the effective time-reversal capability demonstrated here, together with technical improvements to magnetic field stability and homogeneity, our system could provide a fruitful platform to measure out-of-time order correlations and scrambling of quantum information \cite{Garttner2017}
or test spectroscopic protocols that use time reversal to relax the detection resolution required for spectroscopy beyond the standard quantum limit \cite{SchleierSmith2016}. 

\noindent\textbf{{Methods and Materials}}

\noindent
The neutral atomic sample is prepared using laser cooling, magnetic and optical trapping, and evaporative and sympathetic cooling, in an atom-chip apparatus described previously \cite{Bardon:2014}. The hyperfine states used to encode spin information are the $F=9/2$, $m_F = -9/2$ (for $\ket{\downarrow}$) and $m_F = -7/2$ (for $\ket{\uparrow}$) states. At the beginning of a simulation sequence, the fermionic ensemble has a typical temperature of several hundred nanokelvin, with data taken in the range $T\sim 0.3$--$0.5\,E_F/k_B$, where $E_F =(6 N)^{1/3} \hbar \bar \omega$ is the Fermi energy, and $\bar \omega$ is the geometric mean trap frequency. There is no optical lattice: confinement is produced by a crossed-beam 1064-nm optical dipole trap, and the map in Fig.~1 is purely conceptual.

The magnetic field and its gradients are controlled using a combination of microfabricated wires on the atom chip $\approx$\SI{200}{\micro\meter} from the atoms, and macroscopic coils external to the vacuum system. The field is calibrated through rf spectroscopy of the $\ket\downarrow$-to-$\ket\uparrow$ transition. During a typical experimental run, drifts are $\lesssim$\SI{1}{\micro\tesla}.
Magnetic field gradients, which lead to periodic oscillation of the spin clouds (see Supplementary Materials), are measured by displacing the trap centre and repeating rf spectroscopy. In the optimal configuration, gradients are $\{ 13(1), 12(1), 2(5) \}$\,\SI{}{\micro\tesla}/m. The differential displacements $\Delta_i$ resulting from these gradients are small compared to the harmonic oscillator length $a_{i}^2 = \hbar/m \omega_i$, as dimensionless displacements $\Delta_i/a_{i} =\{ 2.1, 0.4, 0.1 \}\times 10^{-2}$.
In this $\Delta_i \ll a_{i}$ regime, the more general XXZ spin model \cite{Koller:2016el} reduces to the Heisenberg model used here.

The effective axial field $h_i$ in the Heisenberg Hamiltonian is the potential-energy differential between the $\ket\uparrow$ and $\ket\downarrow$ states, as sampled by the occupied motional eigenstates. Since one can always subtract a constant potential term, dynamics depend only on the inhomogeneity in $h_i$, quantified through its rms spread $\widetilde h$. There are both magnetic and optical contributions to $h_i$. 
The leading-order magnetic-field contribution is curvature in real space. Direct spectroscopic data can bound this to $\lesssim 500$\,\SI{}{\micro\tesla}/m$^2$, which is compatible with the $200$\,\SI{}{\micro\tesla}/m$^2$ best-fit curvature in the model.
The optical contribution to $\widetilde h$ is tuned using the polarization of the trapping light. For far-detuned light, the differential fractional vector light shift experienced by the atoms is approximately 
$\mathcal P g_F (\delta_\mathrm{FS}/3 \delta)$, where $\mathcal P$ is the polarization of the light, $g_F$ is the $g$-factor of the hyperfine sublevels, $\delta$ is the frequency detuning, and $\delta_\mathrm{FS}$ is the fine-structure splitting of the electronic excited state. For a full range of polarization $\mathcal P\in[-1,1]$ of the ODT beam propagating along the $z$-axis, the resulting vector light shift should modify the trap frequency along $x$ by roughly $\pm 0.06$~Hz, or $\pm 0.015\%$. This control over $\widetilde h$ allows the vertical exploration in the dynamical phase diagram shown in Fig.~2.

The magnetization $2\mathcal S/N$ is determined using repeated measurements and a maximum likelihood estimator, as follows. Each experimental image measures $N_\uparrow,N_\downarrow$ at the end of a Ramsey sequence. 
The fraction of spin-$\uparrow$, $f={N_\uparrow}/N$, relates to the collective spin as $f={1}/{2}+({\mathcal S^Y}/{N})\cos \theta +({\mathcal S^X}/{N})\sin \theta$, where $\theta$ is the phase lag of the second $\pi/2$ pulse. 
However, the typical evolution times (100 ms) exceed the clock coherence time (1.5 ms), so that the accumulated phase $\phi$ randomizes the orientation of the transverse spin, $\mathcal S^Y=\mathcal S \cos \phi,\mathcal S^X=\mathcal S \sin \phi$. 
This yields $f={1}/{2}+({\mathcal{S}}/{N})\cos (\theta')$, where $\theta'=\theta-\phi$ is a random phase. 
To reconstruct the offset $O$ and amplitude $A$ of a Ramsey fringe $F=O+A\cos(\theta')$ from a set of fractions ${f_i}$ acquired in several experimental runs with the same conditions, we assume a probability distribution 
\begin{equation}
 p(f; A,O)=\int_0^\pi \frac{d\theta'}{\sqrt{2\pi}\sigma\pi} \exp\left[\frac{-(A \cos \theta'+O-f)^2}{2\sigma^2}\right] \nonumber
\end{equation}
which is a convolution of the noise-free probability distribution and Gaussian noise with width $\sigma$, calibrated to be $\sigma=0.01$. We construct a log-likelihood function for a set of $n$ fraction measurements, $\ell(A,O; \{f_i\})= \frac{1}{n}\sum_{i=1}^{n}\log\left(p(f_i; A,O)\right)$, from which we numerically compute the maximum-likelihood amplitude $A$ and fringe offset $O$, as well as confidence intervals. Where $O$ falls outside the range $0.50 \pm 0.05$, we discard the data; otherwise, the peak-to-peak amplitude $2A$ is taken as the best estimate of $2\mathcal S/N$.


\nocite{Knap,Falke:2008dq,Loftus:2002iz,Regal:2004kt,Gaebler:2010,Schneider:2012ke,Jordens:2010,Sagi2018,Lepers:2010cv,OHara:2002br}
\nocite{apsrev41Control}
\bibliography{bibTDP}

\begin{thebibliography}{72}%
\makeatletter
\providecommand \@ifxundefined [1]{%
 \@ifx{#1\undefined}
}%
\providecommand \@ifnum [1]{%
 \ifnum #1\expandafter \@firstoftwo
 \else \expandafter \@secondoftwo
 \fi
}%
\providecommand \@ifx [1]{%
 \ifx #1\expandafter \@firstoftwo
 \else \expandafter \@secondoftwo
 \fi
}%
\providecommand \natexlab [1]{#1}%
\providecommand \enquote  [1]{``#1''}%
\providecommand \bibnamefont  [1]{#1}%
\providecommand \bibfnamefont [1]{#1}%
\providecommand \citenamefont [1]{#1}%
\providecommand \href@noop [0]{\@secondoftwo}%
\providecommand \href [0]{\begingroup \@sanitize@url \@href}%
\providecommand \@href[1]{\@@startlink{#1}\@@href}%
\providecommand \@@href[1]{\endgroup#1\@@endlink}%
\providecommand \@sanitize@url [0]{\catcode `\\12\catcode `\$12\catcode
  `\&12\catcode `\#12\catcode `\^12\catcode `\_12\catcode `\%12\relax}%
\providecommand \@@startlink[1]{}%
\providecommand \@@endlink[0]{}%
\providecommand \url  [0]{\begingroup\@sanitize@url \@url }%
\providecommand \@url [1]{\endgroup\@href {#1}{\urlprefix }}%
\providecommand \urlprefix  [0]{URL }%
\providecommand \Eprint [0]{\href }%
\providecommand \doibase [0]{http://dx.doi.org/}%
\providecommand \selectlanguage [0]{\@gobble}%
\providecommand \bibinfo  [0]{\@secondoftwo}%
\providecommand \bibfield  [0]{\@secondoftwo}%
\providecommand \translation [1]{[#1]}%
\providecommand \BibitemOpen [0]{}%
\providecommand \bibitemStop [0]{}%
\providecommand \bibitemNoStop [0]{.\EOS\space}%
\providecommand \EOS [0]{\spacefactor3000\relax}%
\providecommand \BibitemShut  [1]{\csname bibitem#1\endcsname}%
\let\auto@bib@innerbib\@empty
\bibitem [{\citenamefont {Deutsch}\ \emph {et~al.}(2010)\citenamefont
  {Deutsch}, \citenamefont {Ramirez-Martinez}, \citenamefont {Lacro\^ute},
  \citenamefont {Reinhard}, \citenamefont {Schneider}, \citenamefont {Fuchs},
  \citenamefont {Pi\'echon}, \citenamefont {Lalo\"e}, \citenamefont {Reichel},\
  and\ \citenamefont {Rosenbusch}}]{Deutsch:2010ky}%
  \BibitemOpen
  \bibfield  {author} {\bibinfo {author} {\bibfnamefont {C.}~\bibnamefont
  {Deutsch}}, \bibinfo {author} {\bibfnamefont {F.}~\bibnamefont
  {Ramirez-Martinez}}, \bibinfo {author} {\bibfnamefont {C.}~\bibnamefont
  {Lacro\^ute}}, \bibinfo {author} {\bibfnamefont {F.}~\bibnamefont
  {Reinhard}}, \bibinfo {author} {\bibfnamefont {T.}~\bibnamefont {Schneider}},
  \bibinfo {author} {\bibfnamefont {J.~N.}\ \bibnamefont {Fuchs}}, \bibinfo
  {author} {\bibfnamefont {F.}~\bibnamefont {Pi\'echon}}, \bibinfo {author}
  {\bibfnamefont {F.}~\bibnamefont {Lalo\"e}}, \bibinfo {author} {\bibfnamefont
  {J.}~\bibnamefont {Reichel}}, \ and\ \bibinfo {author} {\bibfnamefont
  {P.}~\bibnamefont {Rosenbusch}},\ }\bibfield  {title} {\enquote {\bibinfo
  {title} {Spin self-rephasing and very long coherence times in a trapped
  atomic ensemble},}\ }\href {\doibase 10.1103/PhysRevLett.105.020401}
  {\bibfield  {journal} {\bibinfo  {journal} {Phys. Rev. Lett.}\ }\textbf
  {\bibinfo {volume} {105}},\ \bibinfo {pages} {020401} (\bibinfo {year}
  {2010})}\BibitemShut {NoStop}%
\bibitem [{\citenamefont {Solaro}\ \emph {et~al.}(2016)\citenamefont {Solaro},
  \citenamefont {Bonnin}, \citenamefont {Combes}, \citenamefont {Lopez},
  \citenamefont {Alauze}, \citenamefont {Fuchs}, \citenamefont {Pi{\'e}chon},\
  and\ \citenamefont {Pereira Dos~Santos}}]{Solaro:2016iv}%
  \BibitemOpen
  \bibfield  {author} {\bibinfo {author} {\bibfnamefont {C.}~\bibnamefont
  {Solaro}}, \bibinfo {author} {\bibfnamefont {A.}~\bibnamefont {Bonnin}},
  \bibinfo {author} {\bibfnamefont {F.}~\bibnamefont {Combes}}, \bibinfo
  {author} {\bibfnamefont {M.}~\bibnamefont {Lopez}}, \bibinfo {author}
  {\bibfnamefont {X.}~\bibnamefont {Alauze}}, \bibinfo {author} {\bibfnamefont
  {J.~N.}\ \bibnamefont {Fuchs}}, \bibinfo {author} {\bibfnamefont
  {F.}~\bibnamefont {Pi{\'e}chon}}, \ and\ \bibinfo {author} {\bibfnamefont
  {F.}~\bibnamefont {Pereira Dos~Santos}},\ }\bibfield  {title} {\enquote
  {\bibinfo {title} {Competition between spin echo and spin self-rephasing in a
  trapped atom interferometer},}\ }\href {\doibase
  10.1103/PhysRevLett.117.163003} {\bibfield  {journal} {\bibinfo  {journal}
  {Phys. Rev. Lett.}\ }\textbf {\bibinfo {volume} {117}},\ \bibinfo {pages}
  {163003} (\bibinfo {year} {2016})}\BibitemShut {NoStop}%
\bibitem [{\citenamefont {Pi\'echon}\ \emph {et~al.}(2009)\citenamefont
  {Pi\'echon}, \citenamefont {Fuchs},\ and\ \citenamefont
  {Lalo\"e}}]{Piechon:2009cr}%
  \BibitemOpen
  \bibfield  {author} {\bibinfo {author} {\bibfnamefont {F.}~\bibnamefont
  {Pi\'echon}}, \bibinfo {author} {\bibfnamefont {J.~N.}\ \bibnamefont
  {Fuchs}}, \ and\ \bibinfo {author} {\bibfnamefont {F.}~\bibnamefont
  {Lalo\"e}},\ }\bibfield  {title} {\enquote {\bibinfo {title} {Cumulative
  identical spin rotation effects in collisionless trapped atomic gases},}\
  }\href {\doibase 10.1103/PhysRevLett.102.215301} {\bibfield  {journal}
  {\bibinfo  {journal} {Phys. Rev. Lett.}\ }\textbf {\bibinfo {volume} {102}},\
  \bibinfo {pages} {215301} (\bibinfo {year} {2009})}\BibitemShut {NoStop}%
\bibitem [{\citenamefont {Norcia}\ \emph {et~al.}(2017)\citenamefont {Norcia},
  \citenamefont {Lewis-Swan}, \citenamefont {Cline}, \citenamefont {Zhu},
  \citenamefont {Rey},\ and\ \citenamefont {Thompson}}]{Norcia2017}%
  \BibitemOpen
  \bibfield  {author} {\bibinfo {author} {\bibfnamefont {M.~A.}\ \bibnamefont
  {Norcia}}, \bibinfo {author} {\bibfnamefont {R.~J.}\ \bibnamefont
  {Lewis-Swan}}, \bibinfo {author} {\bibfnamefont {J.~R.~K.}\ \bibnamefont
  {Cline}}, \bibinfo {author} {\bibfnamefont {B.}~\bibnamefont {Zhu}}, \bibinfo
  {author} {\bibfnamefont {A.~M.}\ \bibnamefont {Rey}}, \ and\ \bibinfo
  {author} {\bibfnamefont {J.~K.}\ \bibnamefont {Thompson}},\ }\bibfield
  {title} {\enquote {\bibinfo {title} {Cavity mediated collective spin exchange
  interactions in a strontium superradiant laser},}\ }\href {\doibase
  10.1126/science.aar3102} {\bibfield  {journal} {\bibinfo  {journal}
  {Science}\ }\textbf {\bibinfo {volume} {361}},\ \bibinfo {pages} {259}
  (\bibinfo {year} {2017})}\BibitemShut {NoStop}%
\bibitem [{\citenamefont {Baumann}\ \emph {et~al.}(2010)\citenamefont
  {Baumann}, \citenamefont {Guerlin}, \citenamefont {Brennecke},\ and\
  \citenamefont {Esslinger}}]{Baumann2010}%
  \BibitemOpen
  \bibfield  {author} {\bibinfo {author} {\bibfnamefont {K.}~\bibnamefont
  {Baumann}}, \bibinfo {author} {\bibfnamefont {C.}~\bibnamefont {Guerlin}},
  \bibinfo {author} {\bibfnamefont {F.}~\bibnamefont {Brennecke}}, \ and\
  \bibinfo {author} {\bibfnamefont {T.}~\bibnamefont {Esslinger}},\ }\bibfield
  {title} {\enquote {\bibinfo {title} {{Dicke quantum phase transition with a
  superfluid gas in an optical cavity}},}\ }\href {\doibase
  10.1038/nature09009} {\bibfield  {journal} {\bibinfo  {journal} {Nature}\
  }\textbf {\bibinfo {volume} {464}},\ \bibinfo {pages} {1301} (\bibinfo {year}
  {2010})}\BibitemShut {NoStop}%
\bibitem [{\citenamefont {Klinder}\ \emph {et~al.}(2015)\citenamefont
  {Klinder}, \citenamefont {Ke{\ss}ler}, \citenamefont {Wolke}, \citenamefont
  {Mathey},\ and\ \citenamefont {Hemmerich}}]{Klinder2015}%
  \BibitemOpen
  \bibfield  {author} {\bibinfo {author} {\bibfnamefont {J.}~\bibnamefont
  {Klinder}}, \bibinfo {author} {\bibfnamefont {H.}~\bibnamefont {Ke{\ss}ler}},
  \bibinfo {author} {\bibfnamefont {M.}~\bibnamefont {Wolke}}, \bibinfo
  {author} {\bibfnamefont {L.}~\bibnamefont {Mathey}}, \ and\ \bibinfo {author}
  {\bibfnamefont {A.}~\bibnamefont {Hemmerich}},\ }\bibfield  {title} {\enquote
  {\bibinfo {title} {{Dynamical phase transition in the open Dicke model}},}\
  }\href {\doibase 10.1073/pnas.1417132112} {\bibfield  {journal} {\bibinfo
  {journal} {Proc. Nat. Acad. Sci.}\ }\textbf {\bibinfo {volume} {112}},\
  \bibinfo {pages} {3290} (\bibinfo {year} {2015})}\BibitemShut {NoStop}%
\bibitem [{\citenamefont {Leonard}\ \emph {et~al.}(2017)\citenamefont
  {Leonard}, \citenamefont {Morales}, \citenamefont {Zupancic}, \citenamefont
  {Esslinger},\ and\ \citenamefont {Donner}}]{Leonard2017}%
  \BibitemOpen
  \bibfield  {author} {\bibinfo {author} {\bibfnamefont {J.}~\bibnamefont
  {Leonard}}, \bibinfo {author} {\bibfnamefont {A.}~\bibnamefont {Morales}},
  \bibinfo {author} {\bibfnamefont {P.}~\bibnamefont {Zupancic}}, \bibinfo
  {author} {\bibfnamefont {T.}~\bibnamefont {Esslinger}}, \ and\ \bibinfo
  {author} {\bibfnamefont {T.}~\bibnamefont {Donner}},\ }\bibfield  {title}
  {\enquote {\bibinfo {title} {Supersolid formation in a quantum gas breaking a
  continuous translational symmetry},}\ }\href {\doibase 10.1038/nature21067}
  {\bibfield  {journal} {\bibinfo  {journal} {Nature}\ }\textbf {\bibinfo
  {volume} {543}},\ \bibinfo {pages} {87} (\bibinfo {year} {2017})}\BibitemShut
  {NoStop}%
\bibitem [{\citenamefont {Li}\ \emph {et~al.}(2017)\citenamefont {Li},
  \citenamefont {Jeongwon}, \citenamefont {Huang}, \citenamefont {Shteynas},
  \citenamefont {Cagri}, \citenamefont {Jamison},\ and\ \citenamefont
  {Ketterle}}]{Li:2017}%
  \BibitemOpen
  \bibfield  {author} {\bibinfo {author} {\bibfnamefont {J.}~\bibnamefont
  {Li}}, \bibinfo {author} {\bibfnamefont {L.}~\bibnamefont {Jeongwon}},
  \bibinfo {author} {\bibfnamefont {W.}~\bibnamefont {Huang}}, \bibinfo
  {author} {\bibfnamefont {B.}~\bibnamefont {Shteynas}}, \bibinfo {author}
  {\bibfnamefont {F.}~\bibnamefont {Cagri}}, \bibinfo {author} {\bibfnamefont
  {A.~O.}\ \bibnamefont {Jamison}}, \ and\ \bibinfo {author} {\bibfnamefont
  {W.}~\bibnamefont {Ketterle}},\ }\bibfield  {title} {\enquote {\bibinfo
  {title} {{A stripe phase with supersolid properties in spin orbit-coupled
  Bose-Einstein condensates}},}\ }\href {\doibase 10.1038/nature21431}
  {\bibfield  {journal} {\bibinfo  {journal} {Nature}\ }\textbf {\bibinfo
  {volume} {543}},\ \bibinfo {pages} {91} (\bibinfo {year} {2017})}\BibitemShut
  {NoStop}%
\bibitem [{\citenamefont {Neill}\ \emph {et~al.}(2016)\citenamefont {Neill},
  \citenamefont {Roushan}, \citenamefont {Fang}, \citenamefont {Chen},
  \citenamefont {Kolodrubetz}, \citenamefont {Chen}, \citenamefont {Megrant},
  \citenamefont {Barends}, \citenamefont {Campbell}, \citenamefont {Chiaro},
  \citenamefont {Dunsworth}, \citenamefont {Jeffrey}, \citenamefont {Kelly},
  \citenamefont {Mutus}, \citenamefont {O{\textquoteright}Malley},
  \citenamefont {Quintana}, \citenamefont {Sank}, \citenamefont {Vainsencher},
  \citenamefont {Wenner}, \citenamefont {White}, \citenamefont {Polkovnikov},\
  and\ \citenamefont {Martinis}}]{Neill2016}%
  \BibitemOpen
  \bibfield  {author} {\bibinfo {author} {\bibfnamefont {C.}~\bibnamefont
  {Neill}}, \bibinfo {author} {\bibfnamefont {P.}~\bibnamefont {Roushan}},
  \bibinfo {author} {\bibfnamefont {M.}~\bibnamefont {Fang}}, \bibinfo {author}
  {\bibfnamefont {Y.}~\bibnamefont {Chen}}, \bibinfo {author} {\bibfnamefont
  {M.}~\bibnamefont {Kolodrubetz}}, \bibinfo {author} {\bibfnamefont
  {Z.}~\bibnamefont {Chen}}, \bibinfo {author} {\bibfnamefont {A.}~\bibnamefont
  {Megrant}}, \bibinfo {author} {\bibfnamefont {R.}~\bibnamefont {Barends}},
  \bibinfo {author} {\bibfnamefont {B.}~\bibnamefont {Campbell}}, \bibinfo
  {author} {\bibfnamefont {B.}~\bibnamefont {Chiaro}}, \bibinfo {author}
  {\bibfnamefont {A.}~\bibnamefont {Dunsworth}}, \bibinfo {author}
  {\bibfnamefont {E.}~\bibnamefont {Jeffrey}}, \bibinfo {author} {\bibfnamefont
  {J.}~\bibnamefont {Kelly}}, \bibinfo {author} {\bibfnamefont
  {J.}~\bibnamefont {Mutus}}, \bibinfo {author} {\bibfnamefont {P.~J.~J.}\
  \bibnamefont {O{\textquoteright}Malley}}, \bibinfo {author} {\bibfnamefont
  {C.}~\bibnamefont {Quintana}}, \bibinfo {author} {\bibfnamefont
  {D.}~\bibnamefont {Sank}}, \bibinfo {author} {\bibfnamefont {A.}~\bibnamefont
  {Vainsencher}}, \bibinfo {author} {\bibfnamefont {J.}~\bibnamefont {Wenner}},
  \bibinfo {author} {\bibfnamefont {T.~C.}\ \bibnamefont {White}}, \bibinfo
  {author} {\bibfnamefont {A.}~\bibnamefont {Polkovnikov}}, \ and\ \bibinfo
  {author} {\bibfnamefont {J.~M.}\ \bibnamefont {Martinis}},\ }\bibfield
  {title} {\enquote {\bibinfo {title} {{Ergodic dynamics and thermalization in
  an isolated quantum system}},}\ }\href {\doibase 10.1038/nphys3830}
  {\bibfield  {journal} {\bibinfo  {journal} {Nat Phys}\ }\textbf {\bibinfo
  {volume} {12}},\ \bibinfo {pages} {1037} (\bibinfo {year}
  {2016})}\BibitemShut {NoStop}%
\bibitem [{\citenamefont {Chaudhury}\ \emph {et~al.}(2009)\citenamefont
  {Chaudhury}, \citenamefont {Smith}, \citenamefont {Anderson}, \citenamefont
  {Ghose},\ and\ \citenamefont {Jessen}}]{Chaudhury}%
  \BibitemOpen
  \bibfield  {author} {\bibinfo {author} {\bibfnamefont {S.}~\bibnamefont
  {Chaudhury}}, \bibinfo {author} {\bibfnamefont {A.}~\bibnamefont {Smith}},
  \bibinfo {author} {\bibfnamefont {B.~E.}\ \bibnamefont {Anderson}}, \bibinfo
  {author} {\bibfnamefont {S.}~\bibnamefont {Ghose}}, \ and\ \bibinfo {author}
  {\bibfnamefont {P.~S.}\ \bibnamefont {Jessen}},\ }\bibfield  {title}
  {\enquote {\bibinfo {title} {Quantum signatures of chaos in a kicked top},}\
  }\href {\doibase 10.1038/nature08396
  https://www.nature.com/articles/nature08396#supplementary-information}
  {\bibfield  {journal} {\bibinfo  {journal} {Nature}\ }\textbf {\bibinfo
  {volume} {461}},\ \bibinfo {pages} {768} (\bibinfo {year}
  {2009})}\BibitemShut {NoStop}%
\bibitem [{\citenamefont {Jurcevic}\ \emph {et~al.}(2017)\citenamefont
  {Jurcevic}, \citenamefont {Shen}, \citenamefont {Hauke}, \citenamefont
  {Maier}, \citenamefont {Brydges}, \citenamefont {Hempel}, \citenamefont
  {Lanyon}, \citenamefont {Heyl}, \citenamefont {Blatt},\ and\ \citenamefont
  {Roos}}]{Jurcevic2017}%
  \BibitemOpen
  \bibfield  {author} {\bibinfo {author} {\bibfnamefont {P.}~\bibnamefont
  {Jurcevic}}, \bibinfo {author} {\bibfnamefont {H.}~\bibnamefont {Shen}},
  \bibinfo {author} {\bibfnamefont {P.}~\bibnamefont {Hauke}}, \bibinfo
  {author} {\bibfnamefont {C.}~\bibnamefont {Maier}}, \bibinfo {author}
  {\bibfnamefont {T.}~\bibnamefont {Brydges}}, \bibinfo {author} {\bibfnamefont
  {C.}~\bibnamefont {Hempel}}, \bibinfo {author} {\bibfnamefont {B.~P.}\
  \bibnamefont {Lanyon}}, \bibinfo {author} {\bibfnamefont {M.}~\bibnamefont
  {Heyl}}, \bibinfo {author} {\bibfnamefont {R.}~\bibnamefont {Blatt}}, \ and\
  \bibinfo {author} {\bibfnamefont {C.~F.}\ \bibnamefont {Roos}},\ }\bibfield
  {title} {\enquote {\bibinfo {title} {{Direct observation of dynamical quantum
  phase transitions in an interacting many-body system}},}\ }\href {\doibase
  10.1103/PhysRevLett.119.080501} {\bibfield  {journal} {\bibinfo  {journal}
  {Phys. Rev. Lett.}\ }\textbf {\bibinfo {volume} {119}},\ \bibinfo {pages}
  {080501} (\bibinfo {year} {2017})}\BibitemShut {NoStop}%
\bibitem [{\citenamefont {Zhang}\ \emph
  {et~al.}(2017{\natexlab{a}})\citenamefont {Zhang}, \citenamefont {Hess},
  \citenamefont {Kyprianidis}, \citenamefont {Becker}, \citenamefont {Lee},
  \citenamefont {Smith}, \citenamefont {Pagano}, \citenamefont {Potirniche},
  \citenamefont {Potter}, \citenamefont {Vishwanath}, \citenamefont {Yao},\
  and\ \citenamefont {Monroe}}]{Zhang:2017ci}%
  \BibitemOpen
  \bibfield  {author} {\bibinfo {author} {\bibfnamefont {J.}~\bibnamefont
  {Zhang}}, \bibinfo {author} {\bibfnamefont {P.~W.}\ \bibnamefont {Hess}},
  \bibinfo {author} {\bibfnamefont {A.}~\bibnamefont {Kyprianidis}}, \bibinfo
  {author} {\bibfnamefont {P.}~\bibnamefont {Becker}}, \bibinfo {author}
  {\bibfnamefont {A.}~\bibnamefont {Lee}}, \bibinfo {author} {\bibfnamefont
  {J.}~\bibnamefont {Smith}}, \bibinfo {author} {\bibfnamefont
  {G.}~\bibnamefont {Pagano}}, \bibinfo {author} {\bibfnamefont {I.~D.}\
  \bibnamefont {Potirniche}}, \bibinfo {author} {\bibfnamefont {A.~C.}\
  \bibnamefont {Potter}}, \bibinfo {author} {\bibfnamefont {A.}~\bibnamefont
  {Vishwanath}}, \bibinfo {author} {\bibfnamefont {N.~Y.}\ \bibnamefont {Yao}},
  \ and\ \bibinfo {author} {\bibfnamefont {C.}~\bibnamefont {Monroe}},\
  }\bibfield  {title} {\enquote {\bibinfo {title} {{Observation of a discrete
  time crystal}},}\ }\href {\doibase 10.1038/nature21413} {\bibfield  {journal}
  {\bibinfo  {journal} {Nature}\ }\textbf {\bibinfo {volume} {543}},\ \bibinfo
  {pages} {217} (\bibinfo {year} {2017}{\natexlab{a}})}\BibitemShut {NoStop}%
\bibitem [{\citenamefont {Choi}\ \emph {et~al.}(2017)\citenamefont {Choi},
  \citenamefont {Choi}, \citenamefont {Landig}, \citenamefont {Kucsko},
  \citenamefont {Zhou}, \citenamefont {Isoya}, \citenamefont {Jelezko},
  \citenamefont {Onoda}, \citenamefont {Sumiya}, \citenamefont {Khemani},
  \citenamefont {von Keyserlingk}, \citenamefont {Yao}, \citenamefont
  {Demler},\ and\ \citenamefont {Lukin}}]{Choi2017}%
  \BibitemOpen
  \bibfield  {author} {\bibinfo {author} {\bibfnamefont {S.}~\bibnamefont
  {Choi}}, \bibinfo {author} {\bibfnamefont {J.}~\bibnamefont {Choi}}, \bibinfo
  {author} {\bibfnamefont {R.}~\bibnamefont {Landig}}, \bibinfo {author}
  {\bibfnamefont {G.}~\bibnamefont {Kucsko}}, \bibinfo {author} {\bibfnamefont
  {H.~Y.}\ \bibnamefont {Zhou}}, \bibinfo {author} {\bibfnamefont
  {J.}~\bibnamefont {Isoya}}, \bibinfo {author} {\bibfnamefont
  {F.}~\bibnamefont {Jelezko}}, \bibinfo {author} {\bibfnamefont
  {S.}~\bibnamefont {Onoda}}, \bibinfo {author} {\bibfnamefont
  {H.}~\bibnamefont {Sumiya}}, \bibinfo {author} {\bibfnamefont
  {V.}~\bibnamefont {Khemani}}, \bibinfo {author} {\bibfnamefont
  {C.}~\bibnamefont {von Keyserlingk}}, \bibinfo {author} {\bibfnamefont
  {N.~Y.}\ \bibnamefont {Yao}}, \bibinfo {author} {\bibfnamefont
  {E.}~\bibnamefont {Demler}}, \ and\ \bibinfo {author} {\bibfnamefont {M.~D.}\
  \bibnamefont {Lukin}},\ }\bibfield  {title} {\enquote {\bibinfo {title}
  {Observation of discrete time-crystalline order in a disordered dipolar
  many-body system},}\ }\href {\doibase 10.1038/nature21426} {\bibfield
  {journal} {\bibinfo  {journal} {Nature}\ }\textbf {\bibinfo {volume} {543}},\
  \bibinfo {pages} {221} (\bibinfo {year} {2017})}\BibitemShut {NoStop}%
\bibitem [{\citenamefont {Eckstein}\ \emph {et~al.}(2009)\citenamefont
  {Eckstein}, \citenamefont {Kollar},\ and\ \citenamefont
  {Werner}}]{eckstein09}%
  \BibitemOpen
  \bibfield  {author} {\bibinfo {author} {\bibfnamefont {M.}~\bibnamefont
  {Eckstein}}, \bibinfo {author} {\bibfnamefont {M.}~\bibnamefont {Kollar}}, \
  and\ \bibinfo {author} {\bibfnamefont {P.}~\bibnamefont {Werner}},\
  }\bibfield  {title} {\enquote {\bibinfo {title} {Thermalization after an
  interaction quench in the {H}ubbard model},}\ }\href {\doibase
  10.1103/PhysRevLett.103.056403} {\bibfield  {journal} {\bibinfo  {journal}
  {Phys. Rev. Lett.}\ }\textbf {\bibinfo {volume} {103}},\ \bibinfo {pages}
  {056403} (\bibinfo {year} {2009})}\BibitemShut {NoStop}%
\bibitem [{\citenamefont {Schir\'o}\ and\ \citenamefont
  {Fabrizio}(2010)}]{Schiro10}%
  \BibitemOpen
  \bibfield  {author} {\bibinfo {author} {\bibfnamefont {M.}~\bibnamefont
  {Schir\'o}}\ and\ \bibinfo {author} {\bibfnamefont {M.}~\bibnamefont
  {Fabrizio}},\ }\bibfield  {title} {\enquote {\bibinfo {title} {Time-dependent
  mean field theory for quench dynamics in correlated electron systems},}\
  }\href {\doibase 10.1103/PhysRevLett.105.076401} {\bibfield  {journal}
  {\bibinfo  {journal} {Phys. Rev. Lett.}\ }\textbf {\bibinfo {volume} {105}},\
  \bibinfo {pages} {076401} (\bibinfo {year} {2010})}\BibitemShut {NoStop}%
\bibitem [{\citenamefont {Sciolla}\ and\ \citenamefont
  {Biroli}(2010)}]{Sciolla}%
  \BibitemOpen
  \bibfield  {author} {\bibinfo {author} {\bibfnamefont {B.}~\bibnamefont
  {Sciolla}}\ and\ \bibinfo {author} {\bibfnamefont {G.}~\bibnamefont
  {Biroli}},\ }\bibfield  {title} {\enquote {\bibinfo {title} {Quantum quenches
  and off-equilibrium dynamical transition in the infinite-dimensional
  bose-hubbard model},}\ }\href {\doibase 10.1103/PhysRevLett.105.220401}
  {\bibfield  {journal} {\bibinfo  {journal} {Phys. Rev. Lett.}\ }\textbf
  {\bibinfo {volume} {105}},\ \bibinfo {pages} {220401} (\bibinfo {year}
  {2010})}\BibitemShut {NoStop}%
\bibitem [{\citenamefont {Gambassi}\ and\ \citenamefont
  {Calabrese}(2011)}]{gambassi}%
  \BibitemOpen
  \bibfield  {author} {\bibinfo {author} {\bibfnamefont {A.}~\bibnamefont
  {Gambassi}}\ and\ \bibinfo {author} {\bibfnamefont {P.}~\bibnamefont
  {Calabrese}},\ }\bibfield  {title} {\enquote {\bibinfo {title} {Quantum
  quenches as classical critical films},}\ }\href {\doibase
  10.1209/0295-5075/95/66007} {\bibfield  {journal} {\bibinfo  {journal}
  {Europhys. Lett.}\ }\textbf {\bibinfo {volume} {95}},\ \bibinfo {pages}
  {66007} (\bibinfo {year} {2011})}\BibitemShut {NoStop}%
\bibitem [{\citenamefont {Zhang}\ \emph
  {et~al.}(2017{\natexlab{b}})\citenamefont {Zhang}, \citenamefont {Pagano},
  \citenamefont {Hess}, \citenamefont {Kyprianidis}, \citenamefont {Becker},
  \citenamefont {Kaplan}, \citenamefont {Gorshkov}, \citenamefont {Gong},\ and\
  \citenamefont {Monroe}}]{zhang}%
  \BibitemOpen
  \bibfield  {author} {\bibinfo {author} {\bibfnamefont {J.}~\bibnamefont
  {Zhang}}, \bibinfo {author} {\bibfnamefont {G.}~\bibnamefont {Pagano}},
  \bibinfo {author} {\bibfnamefont {P.~W.}\ \bibnamefont {Hess}}, \bibinfo
  {author} {\bibfnamefont {A.}~\bibnamefont {Kyprianidis}}, \bibinfo {author}
  {\bibfnamefont {P.}~\bibnamefont {Becker}}, \bibinfo {author} {\bibfnamefont
  {H.}~\bibnamefont {Kaplan}}, \bibinfo {author} {\bibfnamefont {A.~V.}\
  \bibnamefont {Gorshkov}}, \bibinfo {author} {\bibfnamefont {Z.~X.}\
  \bibnamefont {Gong}}, \ and\ \bibinfo {author} {\bibfnamefont
  {C.}~\bibnamefont {Monroe}},\ }\bibfield  {title} {\enquote {\bibinfo {title}
  {Observation of a many-body dynamical phase transition with a 53-qubit
  quantum simulator},}\ }\href {\doibase 10.1038/nature24654} {\bibfield
  {journal} {\bibinfo  {journal} {Nature}\ }\textbf {\bibinfo {volume} {551}},\
  \bibinfo {pages} {601} (\bibinfo {year} {2017}{\natexlab{b}})}\BibitemShut
  {NoStop}%
\bibitem [{\citenamefont {Diehl}\ \emph {et~al.}(2010)\citenamefont {Diehl},
  \citenamefont {Tomadin}, \citenamefont {Micheli}, \citenamefont {Fazio},\
  and\ \citenamefont {Zoller}}]{Zoller}%
  \BibitemOpen
  \bibfield  {author} {\bibinfo {author} {\bibfnamefont {S.}~\bibnamefont
  {Diehl}}, \bibinfo {author} {\bibfnamefont {A.}~\bibnamefont {Tomadin}},
  \bibinfo {author} {\bibfnamefont {A.}~\bibnamefont {Micheli}}, \bibinfo
  {author} {\bibfnamefont {R.}~\bibnamefont {Fazio}}, \ and\ \bibinfo {author}
  {\bibfnamefont {P.}~\bibnamefont {Zoller}},\ }\bibfield  {title} {\enquote
  {\bibinfo {title} {Dynamical phase transitions and instabilities in open
  atomic many-body systems},}\ }\href {\doibase 10.1103/PhysRevLett.105.015702}
  {\bibfield  {journal} {\bibinfo  {journal} {Phys. Rev. Lett.}\ }\textbf
  {\bibinfo {volume} {105}},\ \bibinfo {pages} {015702} (\bibinfo {year}
  {2010})}\BibitemShut {NoStop}%
\bibitem [{\citenamefont {Sieberer}\ \emph {et~al.}(2013)\citenamefont
  {Sieberer}, \citenamefont {Huber}, \citenamefont {Altman},\ and\
  \citenamefont {Diehl}}]{Altman}%
  \BibitemOpen
  \bibfield  {author} {\bibinfo {author} {\bibfnamefont {L.~M.}\ \bibnamefont
  {Sieberer}}, \bibinfo {author} {\bibfnamefont {S.~D.}\ \bibnamefont {Huber}},
  \bibinfo {author} {\bibfnamefont {E.}~\bibnamefont {Altman}}, \ and\ \bibinfo
  {author} {\bibfnamefont {S.}~\bibnamefont {Diehl}},\ }\bibfield  {title}
  {\enquote {\bibinfo {title} {Dynamical critical phenomena in
  driven-dissipative systems},}\ }\href {\doibase
  10.1103/PhysRevLett.110.195301} {\bibfield  {journal} {\bibinfo  {journal}
  {Phys. Rev. Lett.}\ }\textbf {\bibinfo {volume} {110}},\ \bibinfo {pages}
  {195301} (\bibinfo {year} {2013})}\BibitemShut {NoStop}%
\bibitem [{\citenamefont {Marino}\ and\ \citenamefont {Diehl}(2016)}]{Marino}%
  \BibitemOpen
  \bibfield  {author} {\bibinfo {author} {\bibfnamefont {J.}~\bibnamefont
  {Marino}}\ and\ \bibinfo {author} {\bibfnamefont {S.}~\bibnamefont {Diehl}},\
  }\bibfield  {title} {\enquote {\bibinfo {title} {Quantum dynamical field
  theory for nonequilibrium phase transitions in driven open systems},}\ }\href
  {\doibase 10.1103/PhysRevB.94.085150} {\bibfield  {journal} {\bibinfo
  {journal} {Phys. Rev. B}\ }\textbf {\bibinfo {volume} {94}},\ \bibinfo
  {pages} {085150} (\bibinfo {year} {2016})}\BibitemShut {NoStop}%
\bibitem [{\citenamefont {Fl{\"a}schner}\ \emph {et~al.}(2018)\citenamefont
  {Fl{\"a}schner}, \citenamefont {Vogel}, \citenamefont {Tarnowski},
  \citenamefont {Rem}, \citenamefont {L{\"u}hmann}, \citenamefont {Heyl},
  \citenamefont {Budich}, \citenamefont {Mathey}, \citenamefont {Sengstock},\
  and\ \citenamefont {Weitenberg}}]{Flaschner:2018ch}%
  \BibitemOpen
  \bibfield  {author} {\bibinfo {author} {\bibfnamefont {N.}~\bibnamefont
  {Fl{\"a}schner}}, \bibinfo {author} {\bibfnamefont {D.}~\bibnamefont
  {Vogel}}, \bibinfo {author} {\bibfnamefont {M.}~\bibnamefont {Tarnowski}},
  \bibinfo {author} {\bibfnamefont {B.~S.}\ \bibnamefont {Rem}}, \bibinfo
  {author} {\bibfnamefont {D.~S.}\ \bibnamefont {L{\"u}hmann}}, \bibinfo
  {author} {\bibfnamefont {M.}~\bibnamefont {Heyl}}, \bibinfo {author}
  {\bibfnamefont {J.~C.}\ \bibnamefont {Budich}}, \bibinfo {author}
  {\bibfnamefont {L.}~\bibnamefont {Mathey}}, \bibinfo {author} {\bibfnamefont
  {K.}~\bibnamefont {Sengstock}}, \ and\ \bibinfo {author} {\bibfnamefont
  {C.}~\bibnamefont {Weitenberg}},\ }\bibfield  {title} {\enquote {\bibinfo
  {title} {{Observation of dynamical vortices after quenches in a system with
  topology}},}\ }\href@noop {} {\bibfield  {journal} {\bibinfo  {journal}
  {Nature Phys.}\ }\textbf {\bibinfo {volume} {14}},\ \bibinfo {pages} {265}
  (\bibinfo {year} {2018})}\BibitemShut {NoStop}%
\bibitem [{\citenamefont {Heyl}\ \emph {et~al.}(2013)\citenamefont {Heyl},
  \citenamefont {Polkovnikov},\ and\ \citenamefont {Kehrein}}]{heyl13}%
  \BibitemOpen
  \bibfield  {author} {\bibinfo {author} {\bibfnamefont {M.}~\bibnamefont
  {Heyl}}, \bibinfo {author} {\bibfnamefont {A.}~\bibnamefont {Polkovnikov}}, \
  and\ \bibinfo {author} {\bibfnamefont {S.}~\bibnamefont {Kehrein}},\
  }\bibfield  {title} {\enquote {\bibinfo {title} {Dynamical quantum phase
  transitions in the transverse-field ising model},}\ }\href {\doibase
  10.1103/PhysRevLett.110.135704} {\bibfield  {journal} {\bibinfo  {journal}
  {Phys. Rev. Lett.}\ }\textbf {\bibinfo {volume} {110}},\ \bibinfo {pages}
  {135704} (\bibinfo {year} {2013})}\BibitemShut {NoStop}%
\bibitem [{\citenamefont {\ifmmode \check{Z}\else
  \v{Z}\fi{}unkovi\ifmmode~\check{c}\else \v{c}\fi{}}\ \emph
  {et~al.}(2016)\citenamefont {\ifmmode \check{Z}\else
  \v{Z}\fi{}unkovi\ifmmode~\check{c}\else \v{c}\fi{}}, \citenamefont {Silva},\
  and\ \citenamefont {Fabrizio}}]{BoyanMF}%
  \BibitemOpen
  \bibfield  {author} {\bibinfo {author} {\bibfnamefont {B.}~\bibnamefont
  {\ifmmode \check{Z}\else \v{Z}\fi{}unkovi\ifmmode~\check{c}\else
  \v{c}\fi{}}}, \bibinfo {author} {\bibfnamefont {A.}~\bibnamefont {Silva}}, \
  and\ \bibinfo {author} {\bibfnamefont {M.}~\bibnamefont {Fabrizio}},\
  }\bibfield  {title} {\enquote {\bibinfo {title} {{Dynamical phase transitions
  and Loschmidt echo in the infinite-range XY model}},}\ }\href {\doibase
  10.1098/rsta.2015.0160} {\bibfield  {journal} {\bibinfo  {journal} {Phil.
  Trans. R. Soc. A}\ }\textbf {\bibinfo {volume} {374}},\ \bibinfo {pages}
  {20150160} (\bibinfo {year} {2016})}\BibitemShut {NoStop}%
\bibitem [{\citenamefont {\ifmmode \check{Z}\else
  \v{Z}\fi{}unkovi\ifmmode~\check{c}\else \v{c}\fi{}}\ \emph
  {et~al.}(2018)\citenamefont {\ifmmode \check{Z}\else
  \v{Z}\fi{}unkovi\ifmmode~\check{c}\else \v{c}\fi{}}, \citenamefont {Heyl},
  \citenamefont {Knap},\ and\ \citenamefont {Silva}}]{Bojan2016b}%
  \BibitemOpen
  \bibfield  {author} {\bibinfo {author} {\bibfnamefont {B.}~\bibnamefont
  {\ifmmode \check{Z}\else \v{Z}\fi{}unkovi\ifmmode~\check{c}\else
  \v{c}\fi{}}}, \bibinfo {author} {\bibfnamefont {M.}~\bibnamefont {Heyl}},
  \bibinfo {author} {\bibfnamefont {M.}~\bibnamefont {Knap}}, \ and\ \bibinfo
  {author} {\bibfnamefont {A.}~\bibnamefont {Silva}},\ }\bibfield  {title}
  {\enquote {\bibinfo {title} {Dynamical quantum phase transitions in spin
  chains with long-range interactions: Merging different concepts of
  nonequilibrium criticality},}\ }\href {\doibase
  10.1103/PhysRevLett.120.130601} {\bibfield  {journal} {\bibinfo  {journal}
  {Phys. Rev. Lett.}\ }\textbf {\bibinfo {volume} {120}},\ \bibinfo {pages}
  {130601} (\bibinfo {year} {2018})}\BibitemShut {NoStop}%
\bibitem [{\citenamefont {Lerose}\ \emph {et~al.}(2019)\citenamefont {Lerose},
  \citenamefont {\ifmmode \check{Z}\else
  \v{Z}\fi{}unkovi\ifmmode~\check{c}\else \v{c}\fi{}}, \citenamefont {Marino},
  \citenamefont {Gambassi},\ and\ \citenamefont {Silva}}]{PhysRevB.99.045128}%
  \BibitemOpen
  \bibfield  {author} {\bibinfo {author} {\bibfnamefont {A.}~\bibnamefont
  {Lerose}}, \bibinfo {author} {\bibfnamefont {B.}~\bibnamefont {\ifmmode
  \check{Z}\else \v{Z}\fi{}unkovi\ifmmode~\check{c}\else \v{c}\fi{}}}, \bibinfo
  {author} {\bibfnamefont {J.}~\bibnamefont {Marino}}, \bibinfo {author}
  {\bibfnamefont {A.}~\bibnamefont {Gambassi}}, \ and\ \bibinfo {author}
  {\bibfnamefont {A.}~\bibnamefont {Silva}},\ }\bibfield  {title} {\enquote
  {\bibinfo {title} {Impact of non-equilibrium fluctuations on pre-thermal
  dynamical phase transitions in long-range interacting spin chains},}\ }\href
  {\doibase 10.1103/PhysRevB.99.045128} {\bibfield  {journal} {\bibinfo
  {journal} {Phys. Rev. B}\ }\textbf {\bibinfo {volume} {99}},\ \bibinfo
  {pages} {045128} (\bibinfo {year} {2019})}\BibitemShut {NoStop}%
\bibitem [{\citenamefont {Dukelsky}\ \emph {et~al.}(2004)\citenamefont
  {Dukelsky}, \citenamefont {Pittel},\ and\ \citenamefont
  {Sierra}}]{Dukelsky2004}%
  \BibitemOpen
  \bibfield  {author} {\bibinfo {author} {\bibfnamefont {J.}~\bibnamefont
  {Dukelsky}}, \bibinfo {author} {\bibfnamefont {S.}~\bibnamefont {Pittel}}, \
  and\ \bibinfo {author} {\bibfnamefont {G.}~\bibnamefont {Sierra}},\
  }\bibfield  {title} {\enquote {\bibinfo {title} {{Colloquium: Exactly
  solvable Richardson-Gaudin models for many-body quantum systems}},}\ }\href
  {\doibase 10.1103/RevModPhys.76.643} {\bibfield  {journal} {\bibinfo
  {journal} {Rev. Mod. Phys.}\ }\textbf {\bibinfo {volume} {76}},\ \bibinfo
  {pages} {643} (\bibinfo {year} {2004})}\BibitemShut {NoStop}%
\bibitem [{\citenamefont {Barankov}\ \emph {et~al.}(2004)\citenamefont
  {Barankov}, \citenamefont {Levitov},\ and\ \citenamefont
  {Spivak}}]{Barank04}%
  \BibitemOpen
  \bibfield  {author} {\bibinfo {author} {\bibfnamefont {R.~A.}\ \bibnamefont
  {Barankov}}, \bibinfo {author} {\bibfnamefont {L.~S.}\ \bibnamefont
  {Levitov}}, \ and\ \bibinfo {author} {\bibfnamefont {B.~Z.}\ \bibnamefont
  {Spivak}},\ }\bibfield  {title} {\enquote {\bibinfo {title} {Collective
  {R}abi oscillations and solitons in a time-dependent {BCS} pairing
  problem},}\ }\href {\doibase 10.1103/PhysRevLett.93.160401} {\bibfield
  {journal} {\bibinfo  {journal} {Phys. Rev. Lett.}\ }\textbf {\bibinfo
  {volume} {93}},\ \bibinfo {pages} {160401} (\bibinfo {year}
  {2004})}\BibitemShut {NoStop}%
\bibitem [{\citenamefont {Yuzbashyan}\ \emph {et~al.}(2015)\citenamefont
  {Yuzbashyan}, \citenamefont {Dzero}, \citenamefont {Gurarie},\ and\
  \citenamefont {Foster}}]{Gurarie2015}%
  \BibitemOpen
  \bibfield  {author} {\bibinfo {author} {\bibfnamefont {E.~A.}\ \bibnamefont
  {Yuzbashyan}}, \bibinfo {author} {\bibfnamefont {M.}~\bibnamefont {Dzero}},
  \bibinfo {author} {\bibfnamefont {V.}~\bibnamefont {Gurarie}}, \ and\
  \bibinfo {author} {\bibfnamefont {M.~S.}\ \bibnamefont {Foster}},\ }\bibfield
   {title} {\enquote {\bibinfo {title} {{Quantum quench phase diagrams of an
  s-wave BCS-BEC condensate}},}\ }\href {\doibase 10.1103/PhysRevA.91.033628}
  {\bibfield  {journal} {\bibinfo  {journal} {Phys. Rev. A}\ }\textbf {\bibinfo
  {volume} {91}},\ \bibinfo {pages} {033628} (\bibinfo {year}
  {2015})}\BibitemShut {NoStop}%
\bibitem [{\citenamefont {Kollath}\ \emph {et~al.}(2007)\citenamefont
  {Kollath}, \citenamefont {L\"auchli},\ and\ \citenamefont
  {Altman}}]{kollath}%
  \BibitemOpen
  \bibfield  {author} {\bibinfo {author} {\bibfnamefont {C.}~\bibnamefont
  {Kollath}}, \bibinfo {author} {\bibfnamefont {A.~M.}\ \bibnamefont
  {L\"auchli}}, \ and\ \bibinfo {author} {\bibfnamefont {E.}~\bibnamefont
  {Altman}},\ }\bibfield  {title} {\enquote {\bibinfo {title} {Quench dynamics
  and nonequilibrium phase diagram of the bose-hubbard model},}\ }\href
  {\doibase 10.1103/PhysRevLett.98.180601} {\bibfield  {journal} {\bibinfo
  {journal} {Phys. Rev. Lett.}\ }\textbf {\bibinfo {volume} {98}},\ \bibinfo
  {pages} {180601} (\bibinfo {year} {2007})}\BibitemShut {NoStop}%
\bibitem [{\citenamefont {Hennig}\ \emph {et~al.}(2016)\citenamefont {Hennig},
  \citenamefont {Neff},\ and\ \citenamefont
  {Fleischmann}}]{PhysRevE.93.032219}%
  \BibitemOpen
  \bibfield  {author} {\bibinfo {author} {\bibfnamefont {H.}~\bibnamefont
  {Hennig}}, \bibinfo {author} {\bibfnamefont {T.}~\bibnamefont {Neff}}, \ and\
  \bibinfo {author} {\bibfnamefont {R.}~\bibnamefont {Fleischmann}},\
  }\bibfield  {title} {\enquote {\bibinfo {title} {{Dynamical phase diagram of
  Gaussian wave packets in optical lattices}},}\ }\href {\doibase
  10.1103/PhysRevE.93.032219} {\bibfield  {journal} {\bibinfo  {journal} {Phys.
  Rev. E}\ }\textbf {\bibinfo {volume} {93}},\ \bibinfo {pages} {032219}
  (\bibinfo {year} {2016})}\BibitemShut {NoStop}%
\bibitem [{\citenamefont {Chiocchetta}\ \emph {et~al.}(2017)\citenamefont
  {Chiocchetta}, \citenamefont {Gambassi}, \citenamefont {Diehl},\ and\
  \citenamefont {Marino}}]{marino2}%
  \BibitemOpen
  \bibfield  {author} {\bibinfo {author} {\bibfnamefont {A.}~\bibnamefont
  {Chiocchetta}}, \bibinfo {author} {\bibfnamefont {A.}~\bibnamefont
  {Gambassi}}, \bibinfo {author} {\bibfnamefont {S.}~\bibnamefont {Diehl}}, \
  and\ \bibinfo {author} {\bibfnamefont {J.}~\bibnamefont {Marino}},\
  }\bibfield  {title} {\enquote {\bibinfo {title} {Dynamical crossovers in
  prethermal critical states},}\ }\href {\doibase
  10.1103/PhysRevLett.118.135701} {\bibfield  {journal} {\bibinfo  {journal}
  {Phys. Rev. Lett.}\ }\textbf {\bibinfo {volume} {118}},\ \bibinfo {pages}
  {135701} (\bibinfo {year} {2017})}\BibitemShut {NoStop}%
\bibitem [{\citenamefont {Anker}\ \emph {et~al.}(2005)\citenamefont {Anker},
  \citenamefont {Albiez}, \citenamefont {Gati}, \citenamefont {Hunsmann},
  \citenamefont {Eiermann}, \citenamefont {Trombettoni},\ and\ \citenamefont
  {Oberthaler}}]{Anker2005}%
  \BibitemOpen
  \bibfield  {author} {\bibinfo {author} {\bibfnamefont {T.}~\bibnamefont
  {Anker}}, \bibinfo {author} {\bibfnamefont {M.}~\bibnamefont {Albiez}},
  \bibinfo {author} {\bibfnamefont {R.}~\bibnamefont {Gati}}, \bibinfo {author}
  {\bibfnamefont {S.}~\bibnamefont {Hunsmann}}, \bibinfo {author}
  {\bibfnamefont {B.}~\bibnamefont {Eiermann}}, \bibinfo {author}
  {\bibfnamefont {A.}~\bibnamefont {Trombettoni}}, \ and\ \bibinfo {author}
  {\bibfnamefont {M.~K.}\ \bibnamefont {Oberthaler}},\ }\bibfield  {title}
  {\enquote {\bibinfo {title} {Nonlinear self-trapping of matter waves in
  periodic potentials},}\ }\href {\doibase 10.1103/PhysRevLett.94.020403}
  {\bibfield  {journal} {\bibinfo  {journal} {Phys. Rev. Lett.}\ }\textbf
  {\bibinfo {volume} {94}},\ \bibinfo {pages} {020403} (\bibinfo {year}
  {2005})}\BibitemShut {NoStop}%
\bibitem [{\citenamefont {Albiez}\ \emph {et~al.}(2005)\citenamefont {Albiez},
  \citenamefont {Gati}, \citenamefont {F\"olling}, \citenamefont {Hunsmann},
  \citenamefont {Cristiani},\ and\ \citenamefont {Oberthaler}}]{Albiez2005}%
  \BibitemOpen
  \bibfield  {author} {\bibinfo {author} {\bibfnamefont {M.}~\bibnamefont
  {Albiez}}, \bibinfo {author} {\bibfnamefont {R.}~\bibnamefont {Gati}},
  \bibinfo {author} {\bibfnamefont {J.}~\bibnamefont {F\"olling}}, \bibinfo
  {author} {\bibfnamefont {S.}~\bibnamefont {Hunsmann}}, \bibinfo {author}
  {\bibfnamefont {M.}~\bibnamefont {Cristiani}}, \ and\ \bibinfo {author}
  {\bibfnamefont {M.~K.}\ \bibnamefont {Oberthaler}},\ }\bibfield  {title}
  {\enquote {\bibinfo {title} {Direct observation of tunneling and nonlinear
  self-trapping in a single bosonic {J}osephson junction},}\ }\href {\doibase
  10.1103/PhysRevLett.95.010402} {\bibfield  {journal} {\bibinfo  {journal}
  {Phys. Rev. Lett.}\ }\textbf {\bibinfo {volume} {95}},\ \bibinfo {pages}
  {010402} (\bibinfo {year} {2005})}\BibitemShut {NoStop}%
\bibitem [{\citenamefont {Levy}\ \emph {et~al.}(2007)\citenamefont {Levy},
  \citenamefont {Lahoud}, \citenamefont {Shomroni},\ and\ \citenamefont
  {Steinhauer}}]{Levy2007}%
  \BibitemOpen
  \bibfield  {author} {\bibinfo {author} {\bibfnamefont {S.}~\bibnamefont
  {Levy}}, \bibinfo {author} {\bibfnamefont {E.}~\bibnamefont {Lahoud}},
  \bibinfo {author} {\bibfnamefont {I.}~\bibnamefont {Shomroni}}, \ and\
  \bibinfo {author} {\bibfnamefont {J.}~\bibnamefont {Steinhauer}},\ }\bibfield
   {title} {\enquote {\bibinfo {title} {The a.c. and d.c. {J}osephson effects
  in a {B}ose-{��E}instein condensate},}\ }\href {\doibase
  10.1038/nature06186} {\bibfield  {journal} {\bibinfo  {journal} {Nature}\
  }\textbf {\bibinfo {volume} {449}},\ \bibinfo {pages} {579} (\bibinfo {year}
  {2007})}\BibitemShut {NoStop}%
\bibitem [{\citenamefont {Abbarchi}\ \emph {et~al.}(2013)\citenamefont
  {Abbarchi}, \citenamefont {Amo}, \citenamefont {Sala}, \citenamefont
  {Solnyshkov}, \citenamefont {Flayac}, \citenamefont {Ferrier}, \citenamefont
  {Sagnes}, \citenamefont {Galopin}, \citenamefont {Lemaître}, \citenamefont
  {Malpuech},\ and\ \citenamefont {Bloch}}]{Abbarchi2013}%
  \BibitemOpen
  \bibfield  {author} {\bibinfo {author} {\bibfnamefont {M.}~\bibnamefont
  {Abbarchi}}, \bibinfo {author} {\bibfnamefont {A.}~\bibnamefont {Amo}},
  \bibinfo {author} {\bibfnamefont {V.~G.}\ \bibnamefont {Sala}}, \bibinfo
  {author} {\bibfnamefont {D.~D.}\ \bibnamefont {Solnyshkov}}, \bibinfo
  {author} {\bibfnamefont {H.}~\bibnamefont {Flayac}}, \bibinfo {author}
  {\bibfnamefont {L.}~\bibnamefont {Ferrier}}, \bibinfo {author} {\bibfnamefont
  {I.}~\bibnamefont {Sagnes}}, \bibinfo {author} {\bibfnamefont
  {E.}~\bibnamefont {Galopin}}, \bibinfo {author} {\bibfnamefont
  {A.}~\bibnamefont {Lemaître}}, \bibinfo {author} {\bibfnamefont
  {G.}~\bibnamefont {Malpuech}}, \ and\ \bibinfo {author} {\bibfnamefont
  {J.}~\bibnamefont {Bloch}},\ }\bibfield  {title} {\enquote {\bibinfo {title}
  {{Macroscopic quantum self-trapping and Josephson oscillations of exciton
  polaritons}},}\ }\href {\doibase 10.1038/nphys2609} {\bibfield  {journal}
  {\bibinfo  {journal} {Nature Phys.}\ }\textbf {\bibinfo {volume} {9}},\
  \bibinfo {pages} {275} (\bibinfo {year} {2013})}\BibitemShut {NoStop}%
\bibitem [{\citenamefont {Reinhard}\ \emph {et~al.}(2013)\citenamefont
  {Reinhard}, \citenamefont {Riou}, \citenamefont {Zundel}, \citenamefont
  {Weiss}, \citenamefont {Li}, \citenamefont {Rey},\ and\ \citenamefont
  {Hipolito}}]{Reinhard2013}%
  \BibitemOpen
  \bibfield  {author} {\bibinfo {author} {\bibfnamefont {A.}~\bibnamefont
  {Reinhard}}, \bibinfo {author} {\bibfnamefont {J.-F.}\ \bibnamefont {Riou}},
  \bibinfo {author} {\bibfnamefont {L.~A.}\ \bibnamefont {Zundel}}, \bibinfo
  {author} {\bibfnamefont {D.~S.}\ \bibnamefont {Weiss}}, \bibinfo {author}
  {\bibfnamefont {S.}~\bibnamefont {Li}}, \bibinfo {author} {\bibfnamefont
  {A.~M.}\ \bibnamefont {Rey}}, \ and\ \bibinfo {author} {\bibfnamefont
  {R.}~\bibnamefont {Hipolito}},\ }\bibfield  {title} {\enquote {\bibinfo
  {title} {Self-trapping in an array of coupled {1D} {B}ose gases},}\ }\href
  {\doibase 10.1103/PhysRevLett.110.033001} {\bibfield  {journal} {\bibinfo
  {journal} {Phys. Rev. Lett.}\ }\textbf {\bibinfo {volume} {110}},\ \bibinfo
  {pages} {033001} (\bibinfo {year} {2013})}\BibitemShut {NoStop}%
\bibitem [{\citenamefont {Martin}\ \emph {et~al.}(2013)\citenamefont {Martin},
  \citenamefont {Bishof}, \citenamefont {Swallows}, \citenamefont {Zhang},
  \citenamefont {Benko}, \citenamefont {Von-Stecher}, \citenamefont {Gorshkov},
  \citenamefont {Rey},\ and\ \citenamefont {Ye}}]{Martin:2013kl}%
  \BibitemOpen
  \bibfield  {author} {\bibinfo {author} {\bibfnamefont {M.~J.}\ \bibnamefont
  {Martin}}, \bibinfo {author} {\bibfnamefont {M.}~\bibnamefont {Bishof}},
  \bibinfo {author} {\bibfnamefont {M.~D.}\ \bibnamefont {Swallows}}, \bibinfo
  {author} {\bibfnamefont {X.}~\bibnamefont {Zhang}}, \bibinfo {author}
  {\bibfnamefont {C.}~\bibnamefont {Benko}}, \bibinfo {author} {\bibfnamefont
  {J.}~\bibnamefont {Von-Stecher}}, \bibinfo {author} {\bibfnamefont {A.~V.}\
  \bibnamefont {Gorshkov}}, \bibinfo {author} {\bibfnamefont {A.~M.}\
  \bibnamefont {Rey}}, \ and\ \bibinfo {author} {\bibfnamefont
  {J.}~\bibnamefont {Ye}},\ }\bibfield  {title} {\enquote {\bibinfo {title} {{A
  quantum many-body spin system in an optical lattice clock}},}\ }\href
  {\doibase 10.1126/science.1236929} {\bibfield  {journal} {\bibinfo  {journal}
  {Science}\ }\textbf {\bibinfo {volume} {341}},\ \bibinfo {pages} {632}
  (\bibinfo {year} {2013})}\BibitemShut {NoStop}%
\bibitem [{\citenamefont {Koller}\ \emph {et~al.}(2015)\citenamefont {Koller},
  \citenamefont {Mundinger}, \citenamefont {Wall},\ and\ \citenamefont
  {Rey}}]{Koller:2015dg}%
  \BibitemOpen
  \bibfield  {author} {\bibinfo {author} {\bibfnamefont {A.~P.}\ \bibnamefont
  {Koller}}, \bibinfo {author} {\bibfnamefont {J.}~\bibnamefont {Mundinger}},
  \bibinfo {author} {\bibfnamefont {M.~L.}\ \bibnamefont {Wall}}, \ and\
  \bibinfo {author} {\bibfnamefont {A.~M.}\ \bibnamefont {Rey}},\ }\bibfield
  {title} {\enquote {\bibinfo {title} {{Demagnetization dynamics of
  noninteracting trapped fermions}},}\ }\href {\doibase
  10.1103/PhysRevA.92.033608} {\bibfield  {journal} {\bibinfo  {journal} {Phys.
  Rev. A}\ }\textbf {\bibinfo {volume} {92}},\ \bibinfo {pages} {033608}
  (\bibinfo {year} {2015})}\BibitemShut {NoStop}%
\bibitem [{\citenamefont {Koller}\ \emph {et~al.}(2016)\citenamefont {Koller},
  \citenamefont {Wall}, \citenamefont {Mundinger},\ and\ \citenamefont
  {Rey}}]{Koller:2016el}%
  \BibitemOpen
  \bibfield  {author} {\bibinfo {author} {\bibfnamefont {A.~P.}\ \bibnamefont
  {Koller}}, \bibinfo {author} {\bibfnamefont {M.~L.}\ \bibnamefont {Wall}},
  \bibinfo {author} {\bibfnamefont {J.}~\bibnamefont {Mundinger}}, \ and\
  \bibinfo {author} {\bibfnamefont {A.~M.}\ \bibnamefont {Rey}},\ }\bibfield
  {title} {\enquote {\bibinfo {title} {Dynamics of interacting fermions in
  spin-dependent potentials},}\ }\href {\doibase
  10.1103/PhysRevLett.117.195302} {\bibfield  {journal} {\bibinfo  {journal}
  {Phys. Rev. Lett.}\ }\textbf {\bibinfo {volume} {117}},\ \bibinfo {pages}
  {195302} (\bibinfo {year} {2016})}\BibitemShut {NoStop}%
\bibitem [{\citenamefont {Auerbach}(1994)}]{Auerbach1994}%
  \BibitemOpen
  \bibfield  {author} {\bibinfo {author} {\bibfnamefont {A.}~\bibnamefont
  {Auerbach}},\ }\href@noop {} {\emph {\bibinfo {title} {Interacting electrons
  and quantum magnetism}}}\ (\bibinfo  {publisher} {Springer-Verlag},\ \bibinfo
  {address} {New York},\ \bibinfo {year} {1994})\BibitemShut {NoStop}%
\bibitem [{\citenamefont {Zhang}\ \emph {et~al.}(2014)\citenamefont {Zhang},
  \citenamefont {Bishof}, \citenamefont {Bromley}, \citenamefont {Kraus},
  \citenamefont {Safronova}, \citenamefont {Zoller}, \citenamefont {Rey},\ and\
  \citenamefont {Ye}}]{Zhang2014}%
  \BibitemOpen
  \bibfield  {author} {\bibinfo {author} {\bibfnamefont {X.}~\bibnamefont
  {Zhang}}, \bibinfo {author} {\bibfnamefont {M.}~\bibnamefont {Bishof}},
  \bibinfo {author} {\bibfnamefont {S.~L.}\ \bibnamefont {Bromley}}, \bibinfo
  {author} {\bibfnamefont {C.~V.}\ \bibnamefont {Kraus}}, \bibinfo {author}
  {\bibfnamefont {M.~S.}\ \bibnamefont {Safronova}}, \bibinfo {author}
  {\bibfnamefont {P.}~\bibnamefont {Zoller}}, \bibinfo {author} {\bibfnamefont
  {A.~M.}\ \bibnamefont {Rey}}, \ and\ \bibinfo {author} {\bibfnamefont
  {J.}~\bibnamefont {Ye}},\ }\bibfield  {title} {\enquote {\bibinfo {title}
  {Spectroscopic observation of {SU(N)}-symmetric interactions in {S}r orbital
  magnetism},}\ }\href {\doibase 10.1126/science.1254978} {\bibfield  {journal}
  {\bibinfo  {journal} {Science}\ }\textbf {\bibinfo {volume} {345}},\ \bibinfo
  {pages} {1467} (\bibinfo {year} {2014})}\BibitemShut {NoStop}%
\bibitem [{\citenamefont {Rey}\ \emph {et~al.}(2014)\citenamefont {Rey},
  \citenamefont {Gorshkov}, \citenamefont {Kraus}, \citenamefont {Martin},
  \citenamefont {Bishof}, \citenamefont {Swallows}, \citenamefont {Zhang},
  \citenamefont {Benko}, \citenamefont {Ye}, \citenamefont {Lemke},\ and\
  \citenamefont {Ludlow}}]{Rey2014}%
  \BibitemOpen
  \bibfield  {author} {\bibinfo {author} {\bibfnamefont {A.~M.}\ \bibnamefont
  {Rey}}, \bibinfo {author} {\bibfnamefont {A.~V.}\ \bibnamefont {Gorshkov}},
  \bibinfo {author} {\bibfnamefont {C.~V.}\ \bibnamefont {Kraus}}, \bibinfo
  {author} {\bibfnamefont {M.~J.}\ \bibnamefont {Martin}}, \bibinfo {author}
  {\bibfnamefont {M.}~\bibnamefont {Bishof}}, \bibinfo {author} {\bibfnamefont
  {M.~D.}\ \bibnamefont {Swallows}}, \bibinfo {author} {\bibfnamefont
  {X.}~\bibnamefont {Zhang}}, \bibinfo {author} {\bibfnamefont
  {C.}~\bibnamefont {Benko}}, \bibinfo {author} {\bibfnamefont
  {J.}~\bibnamefont {Ye}}, \bibinfo {author} {\bibfnamefont {N.~D.}\
  \bibnamefont {Lemke}}, \ and\ \bibinfo {author} {\bibfnamefont {A.~D.}\
  \bibnamefont {Ludlow}},\ }\bibfield  {title} {\enquote {\bibinfo {title}
  {Probing many-body interactions in an optical lattice clock},}\ }\href
  {\doibase 10.1016/j.aop.2013.11.002} {\bibfield  {journal} {\bibinfo
  {journal} {Ann. Phys. (NY)}\ }\textbf {\bibinfo {volume} {340}},\ \bibinfo
  {pages} {311} (\bibinfo {year} {2014})}\BibitemShut {NoStop}%
\bibitem [{\citenamefont {Swallows}\ \emph {et~al.}(2010)\citenamefont
  {Swallows}, \citenamefont {Bishof}, \citenamefont {Lin}, \citenamefont
  {Blatt}, \citenamefont {Martin}, \citenamefont {Rey},\ and\ \citenamefont
  {Ye}}]{Swallows:2010er}%
  \BibitemOpen
  \bibfield  {author} {\bibinfo {author} {\bibfnamefont {M.~D.}\ \bibnamefont
  {Swallows}}, \bibinfo {author} {\bibfnamefont {M.}~\bibnamefont {Bishof}},
  \bibinfo {author} {\bibfnamefont {Y.}~\bibnamefont {Lin}}, \bibinfo {author}
  {\bibfnamefont {S.}~\bibnamefont {Blatt}}, \bibinfo {author} {\bibfnamefont
  {M.~J.}\ \bibnamefont {Martin}}, \bibinfo {author} {\bibfnamefont {A.~M.}\
  \bibnamefont {Rey}}, \ and\ \bibinfo {author} {\bibfnamefont
  {J.}~\bibnamefont {Ye}},\ }\bibfield  {title} {\enquote {\bibinfo {title}
  {{Suppression of collisional shifts in a strongly interacting lattice
  clock}},}\ }\href {\doibase 10.1126/science.1196442} {\bibfield  {journal}
  {\bibinfo  {journal} {Science}\ }\textbf {\bibinfo {volume} {331}},\ \bibinfo
  {pages} {1043} (\bibinfo {year} {2010})}\BibitemShut {NoStop}%
\bibitem [{\citenamefont {{Bromley}}\ \emph {et~al.}(2018)\citenamefont
  {{Bromley}}, \citenamefont {{Kolkowitz}}, \citenamefont {{Bothwell}},
  \citenamefont {{Kedar}}, \citenamefont {{Safavi-Naini}}, \citenamefont
  {{Wall}}, \citenamefont {{Salomon}}, \citenamefont {{Rey}},\ and\
  \citenamefont {{Ye}}}]{2018bromley}%
  \BibitemOpen
  \bibfield  {author} {\bibinfo {author} {\bibfnamefont {S.~L.}\ \bibnamefont
  {{Bromley}}}, \bibinfo {author} {\bibfnamefont {S.}~\bibnamefont
  {{Kolkowitz}}}, \bibinfo {author} {\bibfnamefont {T.}~\bibnamefont
  {{Bothwell}}}, \bibinfo {author} {\bibfnamefont {D.}~\bibnamefont {{Kedar}}},
  \bibinfo {author} {\bibfnamefont {A.}~\bibnamefont {{Safavi-Naini}}},
  \bibinfo {author} {\bibfnamefont {M.~L.}\ \bibnamefont {{Wall}}}, \bibinfo
  {author} {\bibfnamefont {C.}~\bibnamefont {{Salomon}}}, \bibinfo {author}
  {\bibfnamefont {A.~M.}\ \bibnamefont {{Rey}}}, \ and\ \bibinfo {author}
  {\bibfnamefont {J.}~\bibnamefont {{Ye}}},\ }\bibfield  {title} {\enquote
  {\bibinfo {title} {Dynamics of interacting fermions under spin-orbit coupling
  in an optical lattice clock},}\ }\href {\doibase 10.1038/s41567-017-0029-0}
  {\bibfield  {journal} {\bibinfo  {journal} {Nature Phys.}\ }\textbf {\bibinfo
  {volume} {14}},\ \bibinfo {pages} {399} (\bibinfo {year} {2018})}\BibitemShut
  {NoStop}%
\bibitem [{\citenamefont {Matsunaga}\ \emph {et~al.}(2014)\citenamefont
  {Matsunaga}, \citenamefont {Tsuji}, \citenamefont {Fujita}, \citenamefont
  {Sugioka}, \citenamefont {Makise}, \citenamefont {Uzawa}, \citenamefont
  {Terai}, \citenamefont {Wang}, \citenamefont {Aoki},\ and\ \citenamefont
  {Shimano}}]{Matsunaga1145}%
  \BibitemOpen
  \bibfield  {author} {\bibinfo {author} {\bibfnamefont {R.}~\bibnamefont
  {Matsunaga}}, \bibinfo {author} {\bibfnamefont {N.}~\bibnamefont {Tsuji}},
  \bibinfo {author} {\bibfnamefont {H.}~\bibnamefont {Fujita}}, \bibinfo
  {author} {\bibfnamefont {A.}~\bibnamefont {Sugioka}}, \bibinfo {author}
  {\bibfnamefont {K.}~\bibnamefont {Makise}}, \bibinfo {author} {\bibfnamefont
  {Y.}~\bibnamefont {Uzawa}}, \bibinfo {author} {\bibfnamefont
  {H.}~\bibnamefont {Terai}}, \bibinfo {author} {\bibfnamefont
  {Z.}~\bibnamefont {Wang}}, \bibinfo {author} {\bibfnamefont {H.}~\bibnamefont
  {Aoki}}, \ and\ \bibinfo {author} {\bibfnamefont {R.}~\bibnamefont
  {Shimano}},\ }\bibfield  {title} {\enquote {\bibinfo {title} {Light-induced
  collective pseudospin precession resonating with {H}iggs mode in a
  superconductor},}\ }\href {\doibase 10.1126/science.1254697} {\bibfield
  {journal} {\bibinfo  {journal} {Science}\ }\textbf {\bibinfo {volume}
  {345}},\ \bibinfo {pages} {1145} (\bibinfo {year} {2014})}\BibitemShut
  {NoStop}%
\bibitem [{\citenamefont {Chin}\ \emph {et~al.}(2010)\citenamefont {Chin},
  \citenamefont {Grimm}, \citenamefont {Julienne},\ and\ \citenamefont
  {Tiesinga}}]{RevModPhys2010}%
  \BibitemOpen
  \bibfield  {author} {\bibinfo {author} {\bibfnamefont {C.}~\bibnamefont
  {Chin}}, \bibinfo {author} {\bibfnamefont {R.}~\bibnamefont {Grimm}},
  \bibinfo {author} {\bibfnamefont {P.}~\bibnamefont {Julienne}}, \ and\
  \bibinfo {author} {\bibfnamefont {E.}~\bibnamefont {Tiesinga}},\ }\bibfield
  {title} {\enquote {\bibinfo {title} {Feshbach resonances in ultracold
  gases},}\ }\href {\doibase 10.1103/RevModPhys.82.1225} {\bibfield  {journal}
  {\bibinfo  {journal} {Rev. Mod. Phys.}\ }\textbf {\bibinfo {volume} {82}},\
  \bibinfo {pages} {1225} (\bibinfo {year} {2010})}\BibitemShut {NoStop}%
\bibitem [{\citenamefont {Natu}\ and\ \citenamefont
  {Mueller}(2009)}]{Natu:2009gt}%
  \BibitemOpen
  \bibfield  {author} {\bibinfo {author} {\bibfnamefont {S.~S.}\ \bibnamefont
  {Natu}}\ and\ \bibinfo {author} {\bibfnamefont {E.~J.}\ \bibnamefont
  {Mueller}},\ }\bibfield  {title} {\enquote {\bibinfo {title} {Anomalous spin
  segregation in a weakly interacting two-component {F}ermi gas},}\ }\href
  {\doibase 10.1103/PhysRevA.79.051601} {\bibfield  {journal} {\bibinfo
  {journal} {Phys. Rev. A}\ }\textbf {\bibinfo {volume} {79}},\ \bibinfo
  {pages} {051601} (\bibinfo {year} {2009})}\BibitemShut {NoStop}%
\bibitem [{\citenamefont {Du}\ \emph {et~al.}(2009)\citenamefont {Du},
  \citenamefont {Zhang}, \citenamefont {Petricka},\ and\ \citenamefont
  {Thomas}}]{Du:2009hm}%
  \BibitemOpen
  \bibfield  {author} {\bibinfo {author} {\bibfnamefont {X.}~\bibnamefont
  {Du}}, \bibinfo {author} {\bibfnamefont {Y.}~\bibnamefont {Zhang}}, \bibinfo
  {author} {\bibfnamefont {J.}~\bibnamefont {Petricka}}, \ and\ \bibinfo
  {author} {\bibfnamefont {J.~E.}\ \bibnamefont {Thomas}},\ }\bibfield  {title}
  {\enquote {\bibinfo {title} {{Controlling spin current in a trapped {F}ermi
  gas}},}\ }\href {\doibase 10.1103/PhysRevLett.103.010401} {\bibfield
  {journal} {\bibinfo  {journal} {Phys. Rev. Lett.}\ }\textbf {\bibinfo
  {volume} {103}},\ \bibinfo {pages} {010401} (\bibinfo {year}
  {2009})}\BibitemShut {NoStop}%
\bibitem [{\citenamefont {Anderson}(1958)}]{Anderson1958}%
  \BibitemOpen
  \bibfield  {author} {\bibinfo {author} {\bibfnamefont {P.~W.}\ \bibnamefont
  {Anderson}},\ }\bibfield  {title} {\enquote {\bibinfo {title} {Random-phase
  approximation in the theory of superconductivity},}\ }\href {\doibase
  10.1103/PhysRev.112.1900} {\bibfield  {journal} {\bibinfo  {journal} {Phys.
  Rev.}\ }\textbf {\bibinfo {volume} {112}},\ \bibinfo {pages} {1900} (\bibinfo
  {year} {1958})}\BibitemShut {NoStop}%
\bibitem [{\citenamefont {Yuzbashyan}\ and\ \citenamefont {Dzero}(2006)}]{Yuz}%
  \BibitemOpen
  \bibfield  {author} {\bibinfo {author} {\bibfnamefont {E.~A.}\ \bibnamefont
  {Yuzbashyan}}\ and\ \bibinfo {author} {\bibfnamefont {M.}~\bibnamefont
  {Dzero}},\ }\bibfield  {title} {\enquote {\bibinfo {title} {Dynamical
  vanishing of the order parameter in a fermionic condensate},}\ }\href
  {\doibase 10.1103/PhysRevLett.96.230404} {\bibfield  {journal} {\bibinfo
  {journal} {Phys. Rev. Lett.}\ }\textbf {\bibinfo {volume} {96}},\ \bibinfo
  {pages} {230404} (\bibinfo {year} {2006})}\BibitemShut {NoStop}%
\bibitem [{\citenamefont {Yuzbashyan}\ \emph {et~al.}(2006)\citenamefont
  {Yuzbashyan}, \citenamefont {Tsyplyatyev},\ and\ \citenamefont
  {Altshuler}}]{Yuz2}%
  \BibitemOpen
  \bibfield  {author} {\bibinfo {author} {\bibfnamefont {E.~A.}\ \bibnamefont
  {Yuzbashyan}}, \bibinfo {author} {\bibfnamefont {O.}~\bibnamefont
  {Tsyplyatyev}}, \ and\ \bibinfo {author} {\bibfnamefont {B.~L.}\ \bibnamefont
  {Altshuler}},\ }\bibfield  {title} {\enquote {\bibinfo {title} {Relaxation
  and persistent oscillations of the order parameter in fermionic
  condensates},}\ }\href {\doibase 10.1103/PhysRevLett.96.097005} {\bibfield
  {journal} {\bibinfo  {journal} {Phys. Rev. Lett.}\ }\textbf {\bibinfo
  {volume} {96}},\ \bibinfo {pages} {097005} (\bibinfo {year}
  {2006})}\BibitemShut {NoStop}%
\bibitem [{\citenamefont {Widera}\ \emph {et~al.}(2008)\citenamefont {Widera},
  \citenamefont {Trotzky}, \citenamefont {Cheinet}, \citenamefont {F\"olling},
  \citenamefont {Gerbier}, \citenamefont {Bloch}, \citenamefont {Gritsev},
  \citenamefont {Lukin},\ and\ \citenamefont {Demler}}]{Widera2008}%
  \BibitemOpen
  \bibfield  {author} {\bibinfo {author} {\bibfnamefont {A.}~\bibnamefont
  {Widera}}, \bibinfo {author} {\bibfnamefont {S.}~\bibnamefont {Trotzky}},
  \bibinfo {author} {\bibfnamefont {P.}~\bibnamefont {Cheinet}}, \bibinfo
  {author} {\bibfnamefont {S.}~\bibnamefont {F\"olling}}, \bibinfo {author}
  {\bibfnamefont {F.}~\bibnamefont {Gerbier}}, \bibinfo {author} {\bibfnamefont
  {I.}~\bibnamefont {Bloch}}, \bibinfo {author} {\bibfnamefont
  {V.}~\bibnamefont {Gritsev}}, \bibinfo {author} {\bibfnamefont {M.~D.}\
  \bibnamefont {Lukin}}, \ and\ \bibinfo {author} {\bibfnamefont
  {E.}~\bibnamefont {Demler}},\ }\bibfield  {title} {\enquote {\bibinfo {title}
  {Quantum spin dynamics of mode-squeezed {L}uttinger liquids in two-component
  atomic gases},}\ }\href {\doibase 10.1103/PhysRevLett.100.140401} {\bibfield
  {journal} {\bibinfo  {journal} {Phys. Rev. Lett.}\ }\textbf {\bibinfo
  {volume} {100}},\ \bibinfo {pages} {140401} (\bibinfo {year}
  {2008})}\BibitemShut {NoStop}%
\bibitem [{\citenamefont {Lhuillier}\ and\ \citenamefont
  {Lalo\"{e}}(1982)}]{Lhuillier:1982wc}%
  \BibitemOpen
  \bibfield  {author} {\bibinfo {author} {\bibfnamefont {C.}~\bibnamefont
  {Lhuillier}}\ and\ \bibinfo {author} {\bibfnamefont {F.}~\bibnamefont
  {Lalo\"{e}}},\ }\bibfield  {title} {\enquote {\bibinfo {title} {{Transport
  properties in a spin-polarized gas, II}},}\ }\href {\doibase
  10.1051/jphys:01982004302022500} {\bibfield  {journal} {\bibinfo  {journal}
  {J. Phys.-Paris}\ }\textbf {\bibinfo {volume} {43}},\ \bibinfo {pages} {225}
  (\bibinfo {year} {1982})}\BibitemShut {NoStop}%
\bibitem [{\citenamefont {Johnson}\ \emph {et~al.}(1984)\citenamefont
  {Johnson}, \citenamefont {Denker}, \citenamefont {Bigelow}, \citenamefont
  {L\'evy}, \citenamefont {Freed},\ and\ \citenamefont {Lee}}]{Johnson}%
  \BibitemOpen
  \bibfield  {author} {\bibinfo {author} {\bibfnamefont {B.~R.}\ \bibnamefont
  {Johnson}}, \bibinfo {author} {\bibfnamefont {J.~S.}\ \bibnamefont {Denker}},
  \bibinfo {author} {\bibfnamefont {N.}~\bibnamefont {Bigelow}}, \bibinfo
  {author} {\bibfnamefont {L.~P.}\ \bibnamefont {L\'evy}}, \bibinfo {author}
  {\bibfnamefont {J.~H.}\ \bibnamefont {Freed}}, \ and\ \bibinfo {author}
  {\bibfnamefont {D.~M.}\ \bibnamefont {Lee}},\ }\bibfield  {title} {\enquote
  {\bibinfo {title} {Observation of nuclear spin waves in spin-polarized atomic
  hydrogen gas},}\ }\href {\doibase 10.1103/PhysRevLett.52.1508} {\bibfield
  {journal} {\bibinfo  {journal} {Phys. Rev. Lett.}\ }\textbf {\bibinfo
  {volume} {53}},\ \bibinfo {pages} {302} (\bibinfo {year} {1984})}\BibitemShut
  {NoStop}%
\bibitem [{\citenamefont {Gully}\ and\ \citenamefont {Mullin}(1984)}]{Gully}%
  \BibitemOpen
  \bibfield  {author} {\bibinfo {author} {\bibfnamefont {W.~J.}\ \bibnamefont
  {Gully}}\ and\ \bibinfo {author} {\bibfnamefont {W.~J.}\ \bibnamefont
  {Mullin}},\ }\bibfield  {title} {\enquote {\bibinfo {title} {Observation of
  spin rotation effects in polarized $^{3}\mathrm{He}$-$^{4}\mathrm{He}$
  mixtures},}\ }\href {\doibase 10.1103/PhysRevLett.52.1810} {\bibfield
  {journal} {\bibinfo  {journal} {Phys. Rev. Lett.}\ }\textbf {\bibinfo
  {volume} {52}},\ \bibinfo {pages} {1810} (\bibinfo {year}
  {1984})}\BibitemShut {NoStop}%
\bibitem [{\citenamefont {McGuirk}\ \emph {et~al.}(2002)\citenamefont
  {McGuirk}, \citenamefont {Lewandowski}, \citenamefont {Harber}, \citenamefont
  {Nikuni}, \citenamefont {Williams},\ and\ \citenamefont
  {Cornell}}]{McGuirk2002}%
  \BibitemOpen
  \bibfield  {author} {\bibinfo {author} {\bibfnamefont {J.~M.}\ \bibnamefont
  {McGuirk}}, \bibinfo {author} {\bibfnamefont {H.~J.}\ \bibnamefont
  {Lewandowski}}, \bibinfo {author} {\bibfnamefont {D.~M.}\ \bibnamefont
  {Harber}}, \bibinfo {author} {\bibfnamefont {T.}~\bibnamefont {Nikuni}},
  \bibinfo {author} {\bibfnamefont {J.~E.}\ \bibnamefont {Williams}}, \ and\
  \bibinfo {author} {\bibfnamefont {E.~A.}\ \bibnamefont {Cornell}},\
  }\bibfield  {title} {\enquote {\bibinfo {title} {Spatial resolution of spin
  waves in an ultracold gas},}\ }\href {\doibase 10.1103/PhysRevLett.89.090402}
  {\bibfield  {journal} {\bibinfo  {journal} {Phys. Rev. Lett.}\ }\textbf
  {\bibinfo {volume} {89}},\ \bibinfo {pages} {090402} (\bibinfo {year}
  {2002})}\BibitemShut {NoStop}%
\bibitem [{\citenamefont {Koschorreck}\ \emph {et~al.}(2013)\citenamefont
  {Koschorreck}, \citenamefont {Pertot}, \citenamefont {Vogt},\ and\
  \citenamefont {K\"{o}hl}}]{Kohl2013}%
  \BibitemOpen
  \bibfield  {author} {\bibinfo {author} {\bibfnamefont {M.}~\bibnamefont
  {Koschorreck}}, \bibinfo {author} {\bibfnamefont {D.}~\bibnamefont {Pertot}},
  \bibinfo {author} {\bibfnamefont {E.}~\bibnamefont {Vogt}}, \ and\ \bibinfo
  {author} {\bibfnamefont {M.}~\bibnamefont {K\"{o}hl}},\ }\bibfield  {title}
  {\enquote {\bibinfo {title} {{Universal spin dynamics in two-dimensional
  {F}ermi gases}},}\ }\href {\doibase doi:10.1038/nphys2637} {\bibfield
  {journal} {\bibinfo  {journal} {Nature Phys.}\ }\textbf {\bibinfo {volume}
  {9}},\ \bibinfo {pages} {405} (\bibinfo {year} {2013})}\BibitemShut {NoStop}%
\bibitem [{\citenamefont {Campbell}\ \emph {et~al.}(2017)\citenamefont
  {Campbell}, \citenamefont {Hutson}, \citenamefont {Marti}, \citenamefont
  {Goban}, \citenamefont {Darkwah~Oppong}, \citenamefont {McNally},
  \citenamefont {Sonderhouse}, \citenamefont {Robinson}, \citenamefont {Zhang},
  \citenamefont {Bloom},\ and\ \citenamefont {Ye}}]{Campbell}%
  \BibitemOpen
  \bibfield  {author} {\bibinfo {author} {\bibfnamefont {S.~L.}\ \bibnamefont
  {Campbell}}, \bibinfo {author} {\bibfnamefont {R.~B.}\ \bibnamefont
  {Hutson}}, \bibinfo {author} {\bibfnamefont {G.~E.}\ \bibnamefont {Marti}},
  \bibinfo {author} {\bibfnamefont {A.}~\bibnamefont {Goban}}, \bibinfo
  {author} {\bibfnamefont {N.}~\bibnamefont {Darkwah~Oppong}}, \bibinfo
  {author} {\bibfnamefont {R.~L.}\ \bibnamefont {McNally}}, \bibinfo {author}
  {\bibfnamefont {L.}~\bibnamefont {Sonderhouse}}, \bibinfo {author}
  {\bibfnamefont {J.~M.}\ \bibnamefont {Robinson}}, \bibinfo {author}
  {\bibfnamefont {W.}~\bibnamefont {Zhang}}, \bibinfo {author} {\bibfnamefont
  {B.~J.}\ \bibnamefont {Bloom}}, \ and\ \bibinfo {author} {\bibfnamefont
  {J.}~\bibnamefont {Ye}},\ }\bibfield  {title} {\enquote {\bibinfo {title} {{A
  Fermi-degenerate three-dimensional optical lattice clock}},}\ }\href
  {\doibase 10.1126/science.aam5538} {\bibfield  {journal} {\bibinfo  {journal}
  {Science}\ }\textbf {\bibinfo {volume} {358}},\ \bibinfo {pages} {90}
  (\bibinfo {year} {2017})}\BibitemShut {NoStop}%
\bibitem [{\citenamefont {G\"arttner}\ \emph {et~al.}(2017)\citenamefont
  {G\"arttner}, \citenamefont {Bohnet}, \citenamefont {Safavi-Naini},
  \citenamefont {Wall}, \citenamefont {Bollinger},\ and\ \citenamefont
  {Rey}}]{Garttner2017}%
  \BibitemOpen
  \bibfield  {author} {\bibinfo {author} {\bibfnamefont {M.}~\bibnamefont
  {G\"arttner}}, \bibinfo {author} {\bibfnamefont {J.~G.}\ \bibnamefont
  {Bohnet}}, \bibinfo {author} {\bibfnamefont {A.}~\bibnamefont
  {Safavi-Naini}}, \bibinfo {author} {\bibfnamefont {M.~L.}\ \bibnamefont
  {Wall}}, \bibinfo {author} {\bibfnamefont {J.~J.}\ \bibnamefont {Bollinger}},
  \ and\ \bibinfo {author} {\bibfnamefont {A.~M.}\ \bibnamefont {Rey}},\
  }\bibfield  {title} {\enquote {\bibinfo {title} {Measuring out-of-time-order
  correlations and multiple quantum spectra in a trapped-ion quantum magnet},}\
  }\href {\doibase 10.1038/nphys4119} {\bibfield  {journal} {\bibinfo
  {journal} {Nature Phys.}\ }\textbf {\bibinfo {volume} {13}},\ \bibinfo
  {pages} {781} (\bibinfo {year} {2017})}\BibitemShut {NoStop}%
\bibitem [{\citenamefont {Davis}\ \emph {et~al.}(2016)\citenamefont {Davis},
  \citenamefont {Bentsen},\ and\ \citenamefont
  {Schleier-Smith}}]{SchleierSmith2016}%
  \BibitemOpen
  \bibfield  {author} {\bibinfo {author} {\bibfnamefont {E.}~\bibnamefont
  {Davis}}, \bibinfo {author} {\bibfnamefont {G.}~\bibnamefont {Bentsen}}, \
  and\ \bibinfo {author} {\bibfnamefont {M.}~\bibnamefont {Schleier-Smith}},\
  }\bibfield  {title} {\enquote {\bibinfo {title} {Approaching the {H}eisenberg
  limit without single-particle detection},}\ }\href {\doibase
  10.1103/PhysRevLett.116.053601} {\bibfield  {journal} {\bibinfo  {journal}
  {Phys. Rev. Lett.}\ }\textbf {\bibinfo {volume} {116}},\ \bibinfo {pages}
  {053601} (\bibinfo {year} {2016})}\BibitemShut {NoStop}%
\bibitem [{\citenamefont {Bardon}\ \emph {et~al.}(2014)\citenamefont {Bardon},
  \citenamefont {Beattie}, \citenamefont {Luciuk}, \citenamefont {Cairncross},
  \citenamefont {Fine}, \citenamefont {Cheng}, \citenamefont {Edge},
  \citenamefont {Taylor}, \citenamefont {Zhang}, \citenamefont {Trotzky},\ and\
  \citenamefont {Thywissen}}]{Bardon:2014}%
  \BibitemOpen
  \bibfield  {author} {\bibinfo {author} {\bibfnamefont {A.~B.}\ \bibnamefont
  {Bardon}}, \bibinfo {author} {\bibfnamefont {S.}~\bibnamefont {Beattie}},
  \bibinfo {author} {\bibfnamefont {C.}~\bibnamefont {Luciuk}}, \bibinfo
  {author} {\bibfnamefont {W.}~\bibnamefont {Cairncross}}, \bibinfo {author}
  {\bibfnamefont {D.}~\bibnamefont {Fine}}, \bibinfo {author} {\bibfnamefont
  {N.~S.}\ \bibnamefont {Cheng}}, \bibinfo {author} {\bibfnamefont {G.~J.~A.}\
  \bibnamefont {Edge}}, \bibinfo {author} {\bibfnamefont {E.}~\bibnamefont
  {Taylor}}, \bibinfo {author} {\bibfnamefont {S.}~\bibnamefont {Zhang}},
  \bibinfo {author} {\bibfnamefont {S.}~\bibnamefont {Trotzky}}, \ and\
  \bibinfo {author} {\bibfnamefont {J.~H.}\ \bibnamefont {Thywissen}},\
  }\bibfield  {title} {\enquote {\bibinfo {title} {{Transverse demagnetization
  dynamics of a unitary Fermi gas}},}\ }\href {\doibase
  10.1126/science.1247425} {\bibfield  {journal} {\bibinfo  {journal}
  {Science}\ }\textbf {\bibinfo {volume} {344}},\ \bibinfo {pages} {722}
  (\bibinfo {year} {2014})}\BibitemShut {NoStop}%
\bibitem [{\citenamefont {Smacchia}\ \emph {et~al.}(2015)\citenamefont
  {Smacchia}, \citenamefont {Knap}, \citenamefont {Demler},\ and\ \citenamefont
  {Silva}}]{Knap}%
  \BibitemOpen
  \bibfield  {author} {\bibinfo {author} {\bibfnamefont {P.}~\bibnamefont
  {Smacchia}}, \bibinfo {author} {\bibfnamefont {M.}~\bibnamefont {Knap}},
  \bibinfo {author} {\bibfnamefont {E.}~\bibnamefont {Demler}}, \ and\ \bibinfo
  {author} {\bibfnamefont {A.}~\bibnamefont {Silva}},\ }\bibfield  {title}
  {\enquote {\bibinfo {title} {Exploring dynamical phase transitions and
  prethermalization with quantum noise of excitations},}\ }\href {\doibase
  10.1103/PhysRevB.91.205136} {\bibfield  {journal} {\bibinfo  {journal} {Phys.
  Rev. B}\ }\textbf {\bibinfo {volume} {91}},\ \bibinfo {pages} {205136}
  (\bibinfo {year} {2015})}\BibitemShut {NoStop}%
\bibitem [{\citenamefont {Falke}\ \emph {et~al.}(2008)\citenamefont {Falke},
  \citenamefont {Knoeckel}, \citenamefont {Friebe}, \citenamefont {Riedmann},
  \citenamefont {Tiemann},\ and\ \citenamefont {Lisdat}}]{Falke:2008dq}%
  \BibitemOpen
  \bibfield  {author} {\bibinfo {author} {\bibfnamefont {S.}~\bibnamefont
  {Falke}}, \bibinfo {author} {\bibfnamefont {H.}~\bibnamefont {Knoeckel}},
  \bibinfo {author} {\bibfnamefont {J.}~\bibnamefont {Friebe}}, \bibinfo
  {author} {\bibfnamefont {M.}~\bibnamefont {Riedmann}}, \bibinfo {author}
  {\bibfnamefont {E.}~\bibnamefont {Tiemann}}, \ and\ \bibinfo {author}
  {\bibfnamefont {C.}~\bibnamefont {Lisdat}},\ }\bibfield  {title} {\enquote
  {\bibinfo {title} {{Potassium ground-state scattering parameters and
  Born-Oppenheimer potentials from molecular spectroscopy}},}\ }\href {\doibase
  10.1103/PhysRevA.78.012503} {\bibfield  {journal} {\bibinfo  {journal} {Phys.
  Rev. A}\ }\textbf {\bibinfo {volume} {78}},\ \bibinfo {pages} {012503}
  (\bibinfo {year} {2008})}\BibitemShut {NoStop}%
\bibitem [{\citenamefont {Loftus}\ \emph {et~al.}(2002)\citenamefont {Loftus},
  \citenamefont {Regal}, \citenamefont {Ticknor}, \citenamefont {Bohn},\ and\
  \citenamefont {Jin}}]{Loftus:2002iz}%
  \BibitemOpen
  \bibfield  {author} {\bibinfo {author} {\bibfnamefont {T.}~\bibnamefont
  {Loftus}}, \bibinfo {author} {\bibfnamefont {C.~A.}\ \bibnamefont {Regal}},
  \bibinfo {author} {\bibfnamefont {C.}~\bibnamefont {Ticknor}}, \bibinfo
  {author} {\bibfnamefont {J.~L.}\ \bibnamefont {Bohn}}, \ and\ \bibinfo
  {author} {\bibfnamefont {D.~S.}\ \bibnamefont {Jin}},\ }\bibfield  {title}
  {\enquote {\bibinfo {title} {{Resonant control of elastic collisions in an
  optically trapped Fermi gas of atoms}},}\ }\href {\doibase
  10.1103/PhysRevLett.88.173201} {\bibfield  {journal} {\bibinfo  {journal}
  {Phys. Rev. Lett.}\ }\textbf {\bibinfo {volume} {88}},\ \bibinfo {pages}
  {173201} (\bibinfo {year} {2002})}\BibitemShut {NoStop}%
\bibitem [{\citenamefont {Regal}\ \emph {et~al.}(2004)\citenamefont {Regal},
  \citenamefont {Greiner},\ and\ \citenamefont {Jin}}]{Regal:2004kt}%
  \BibitemOpen
  \bibfield  {author} {\bibinfo {author} {\bibfnamefont {C.~A.}\ \bibnamefont
  {Regal}}, \bibinfo {author} {\bibfnamefont {M.}~\bibnamefont {Greiner}}, \
  and\ \bibinfo {author} {\bibfnamefont {D.~S.}\ \bibnamefont {Jin}},\
  }\bibfield  {title} {\enquote {\bibinfo {title} {{Observation of resonance
  condensation of fermionic atom pairs}},}\ }\href@noop {} {\bibfield
  {journal} {\bibinfo  {journal} {Phys. Rev. Lett.}\ }\textbf {\bibinfo
  {volume} {92}},\ \bibinfo {pages} {040403} (\bibinfo {year}
  {2004})}\BibitemShut {NoStop}%
\bibitem [{\citenamefont {{Gaebler}}\ \emph {et~al.}(2010)\citenamefont
  {{Gaebler}}, \citenamefont {{Stewart}}, \citenamefont {{Drake}},
  \citenamefont {{Jin}}, \citenamefont {{Perali}}, \citenamefont {{Pieri}},\
  and\ \citenamefont {{Strinati}}}]{Gaebler:2010}%
  \BibitemOpen
  \bibfield  {author} {\bibinfo {author} {\bibfnamefont {J.~P.}\ \bibnamefont
  {{Gaebler}}}, \bibinfo {author} {\bibfnamefont {J.~T.}\ \bibnamefont
  {{Stewart}}}, \bibinfo {author} {\bibfnamefont {T.~E.}\ \bibnamefont
  {{Drake}}}, \bibinfo {author} {\bibfnamefont {D.~S.}\ \bibnamefont {{Jin}}},
  \bibinfo {author} {\bibfnamefont {A.}~\bibnamefont {{Perali}}}, \bibinfo
  {author} {\bibfnamefont {P.}~\bibnamefont {{Pieri}}}, \ and\ \bibinfo
  {author} {\bibfnamefont {G.~C.}\ \bibnamefont {{Strinati}}},\ }\bibfield
  {title} {\enquote {\bibinfo {title} {{Observation of pseudogap behaviour in a
  strongly interacting Fermi gas}},}\ }\href {\doibase 10.1038/nphys1709}
  {\bibfield  {journal} {\bibinfo  {journal} {Nature Phys.}\ }\textbf {\bibinfo
  {volume} {6}},\ \bibinfo {pages} {569} (\bibinfo {year} {2010})}\BibitemShut
  {NoStop}%
\bibitem [{\citenamefont {Schneider}\ \emph {et~al.}(2012)\citenamefont
  {Schneider}, \citenamefont {Hackerm{\"u}ller}, \citenamefont {Ronzheimer},
  \citenamefont {Will}, \citenamefont {Braun}, \citenamefont {Best},
  \citenamefont {Bloch}, \citenamefont {Demler}, \citenamefont {Mandt},
  \citenamefont {Rasch},\ and\ \citenamefont {Rosch}}]{Schneider:2012ke}%
  \BibitemOpen
  \bibfield  {author} {\bibinfo {author} {\bibfnamefont {U.}~\bibnamefont
  {Schneider}}, \bibinfo {author} {\bibfnamefont {L.}~\bibnamefont
  {Hackerm{\"u}ller}}, \bibinfo {author} {\bibfnamefont {J.~P.}\ \bibnamefont
  {Ronzheimer}}, \bibinfo {author} {\bibfnamefont {S.}~\bibnamefont {Will}},
  \bibinfo {author} {\bibfnamefont {S.}~\bibnamefont {Braun}}, \bibinfo
  {author} {\bibfnamefont {T.}~\bibnamefont {Best}}, \bibinfo {author}
  {\bibfnamefont {I.}~\bibnamefont {Bloch}}, \bibinfo {author} {\bibfnamefont
  {E.}~\bibnamefont {Demler}}, \bibinfo {author} {\bibfnamefont
  {S.}~\bibnamefont {Mandt}}, \bibinfo {author} {\bibfnamefont
  {D.}~\bibnamefont {Rasch}}, \ and\ \bibinfo {author} {\bibfnamefont
  {A.}~\bibnamefont {Rosch}},\ }\bibfield  {title} {\enquote {\bibinfo {title}
  {{Fermionic transport and out-of-equilibrium dynamics in a homogeneous
  Hubbard model~with~ultracold atoms}},}\ }\href {\doibase 10.1038/nphys2205}
  {\bibfield  {journal} {\bibinfo  {journal} {Nature Phys.}\ }\textbf {\bibinfo
  {volume} {8}},\ \bibinfo {pages} {213} (\bibinfo {year} {2012})}\BibitemShut
  {NoStop}%
\bibitem [{\citenamefont {J\"ordens}(2010)}]{Jordens:2010}%
  \BibitemOpen
  \bibfield  {author} {\bibinfo {author} {\bibfnamefont {J.}~\bibnamefont
  {J\"ordens}},\ }\emph {\bibinfo {title} {{Metallic and Mott-insulating phases
  in fermionic quantum gases}}},\ \href {\doibase 10.3929/ethz-a-006278918}
  {Ph.D. thesis},\ \bibinfo  {school} {ETH Z\"urich} (\bibinfo {year}
  {2010})\BibitemShut {NoStop}%
\bibitem [{\citenamefont {Shkedrov}\ \emph {et~al.}(2018)\citenamefont
  {Shkedrov}, \citenamefont {Florshaim}, \citenamefont {Ness}, \citenamefont
  {Gandman},\ and\ \citenamefont {Sagi}}]{Sagi2018}%
  \BibitemOpen
  \bibfield  {author} {\bibinfo {author} {\bibfnamefont {C.}~\bibnamefont
  {Shkedrov}}, \bibinfo {author} {\bibfnamefont {Y.}~\bibnamefont {Florshaim}},
  \bibinfo {author} {\bibfnamefont {G.}~\bibnamefont {Ness}}, \bibinfo {author}
  {\bibfnamefont {A.}~\bibnamefont {Gandman}}, \ and\ \bibinfo {author}
  {\bibfnamefont {Y.}~\bibnamefont {Sagi}},\ }\bibfield  {title} {\enquote
  {\bibinfo {title} {High-sensitivity rf spectroscopy of a strongly interacting
  fermi gas},}\ }\href {\doibase 10.1103/PhysRevLett.121.093402} {\bibfield
  {journal} {\bibinfo  {journal} {Phys. Rev. Lett.}\ }\textbf {\bibinfo
  {volume} {121}},\ \bibinfo {pages} {093402} (\bibinfo {year}
  {2018})}\BibitemShut {NoStop}%
\bibitem [{\citenamefont {Lepers}\ \emph {et~al.}(2010)\citenamefont {Lepers},
  \citenamefont {Davesne}, \citenamefont {Chiacchiera},\ and\ \citenamefont
  {Urban}}]{Lepers:2010cv}%
  \BibitemOpen
  \bibfield  {author} {\bibinfo {author} {\bibfnamefont {T.}~\bibnamefont
  {Lepers}}, \bibinfo {author} {\bibfnamefont {D.}~\bibnamefont {Davesne}},
  \bibinfo {author} {\bibfnamefont {S.}~\bibnamefont {Chiacchiera}}, \ and\
  \bibinfo {author} {\bibfnamefont {M.}~\bibnamefont {Urban}},\ }\bibfield
  {title} {\enquote {\bibinfo {title} {{Numerical solution of the Boltzmann
  equation for the collective modes of trapped Fermi gases}},}\ }\href
  {\doibase 10.1103/PhysRevA.82.023609} {\bibfield  {journal} {\bibinfo
  {journal} {Phys. Rev. A}\ }\textbf {\bibinfo {volume} {82}},\ \bibinfo
  {pages} {023609} (\bibinfo {year} {2010})}\BibitemShut {NoStop}%
\bibitem [{\citenamefont {O'Hara}\ \emph {et~al.}(2002)\citenamefont {O'Hara},
  \citenamefont {Hemmer}, \citenamefont {Gehm}, \citenamefont {Granade},\ and\
  \citenamefont {Thomas}}]{OHara:2002br}%
  \BibitemOpen
  \bibfield  {author} {\bibinfo {author} {\bibfnamefont {K.}~\bibnamefont
  {O'Hara}}, \bibinfo {author} {\bibfnamefont {S.}~\bibnamefont {Hemmer}},
  \bibinfo {author} {\bibfnamefont {M.}~\bibnamefont {Gehm}}, \bibinfo {author}
  {\bibfnamefont {S.}~\bibnamefont {Granade}}, \ and\ \bibinfo {author}
  {\bibfnamefont {J.~E.}\ \bibnamefont {Thomas}},\ }\bibfield  {title}
  {\enquote {\bibinfo {title} {{Observation of a strongly interacting
  degenerate Fermi gas of atoms}},}\ }\href {\doibase 10.1126/science.1079107}
  {\bibfield  {journal} {\bibinfo  {journal} {Science}\ }\textbf {\bibinfo
  {volume} {298}},\ \bibinfo {pages} {2179} (\bibinfo {year}
  {2002})}\BibitemShut {NoStop}%
\end{thebibliography}%

\noindent {\bf\small Acknowledgements:} We thank A.~Koller and C.~Luciuk for early work on this project, and V.~Gurarie, B.~Lev, J.~Thompson, M.~Foster, and D.~Stamper-Kurn for discussions. 
{\bf\small Funding:}
This work is supported by NSERC, by the Air Force Office of Scientific Research grants FA9550-13-1-0063, FA9550-18-1-0319 and its Multidisciplinary University Research Initiative grant(MURI), by the Army Research Office grant ARO W911NF-15-1-0603, the Defense Advanced Research Projects Agency (DARPA) and Army Research Office grant W911NF-16-1-0576, the National Science Foundation grant PHY1820885, JILA-NSF grant PFC-173400, and the National Institute of Standards and Technology. 
{\bf\small Author Contributions:} 
The work was conceived by A.R., J.T.\ and S.T.\ 
Experiments were performed by S.S., B.O., H.S., K.J., and S.T.\ 
Data was analyzed by S.S., P.H., and B.O.\ 
Theoretical models and simulation was done by P.H., J.M., J.T., and A.R.
All authors contributed to manuscript preparation. 
{\bf\small Competing Interests:}
The authors declare that they have no competing interests.
{\bf\small Data and materials availability:} 
The datasets generated and analyzed during the current study are available from the corresponding authors upon reasonable request. 

{\small \noindent\textbf{{Supplementary Materials:}}\\
\noindent Fig.~S1.\ Order parameters predicted by Lax vector analysis in a 1D system.\\
Fig.~S2.\ Approximate form of the Lax vector in a 3D system.\\
Fig.~S3.\ Effective mean field potential.\\
Fig.~S4.\ Magnetization dynamics for non-interacting particles.\\
Table~S1.\ Determinations of the $^{40}$K Feshbach resonance parameters.}\\
Fig.~S5.\ Determination of the Feshbach zero-crossing.\\
Fig.~S6.\ Spin-echo amplitude near the Feshbach zero-crossing.\\
Fig.~S7.\ Fits to time series.\\
Fig.~S8.\ Equilibrium scattering rate versus temperature.\\
Fig.~S9.\ Non-equilibrium scattering rate.\\
Refs.~63-72

\end{document}